\documentclass[12pt]{article}

\usepackage{graphicx}
\usepackage{amsmath}
\usepackage {float}
\usepackage{enumerate}

\begin{document}

\begin{center}
{\Large {\bf Rome teleportation experiment analysed in the Wigner representation: the role of the zeropoint fluctuations in complete one-photon polarization-momentum Bell-state analysis}}

\vspace{0.75cm}

{\bf A. Casado$^1$, S. Guerra$^2$, and J. Pl\'{a}cido$^2$}.

$^1$ Departamento de F\'{\i}sica Aplicada III, Escuela T\'ecnica Superior de
Ingenier\'ia, Universidad de Sevilla, 41092 Sevilla, Spain.

Electronic address: acasado@us.es

$^2$ Grupo de Ingenier\'{\i}a T\'ermica e Instrumentaci\'on, Universidad de Las Palmas de Gran
Canaria,

 35017 Las Palmas de Gran Canaria, Spain.
\vspace{1cm}

\end{center}

\vspace{0.5cm}
\noindent{\bf Acknowledgements}

The authors would like to thank Prof. E. Santos for revising the manus\-cript, and for helpful suggestions and comments on the work. We are grateful for the insights gained in conversations with R. Risco-Delgado.
\vspace{0.5cm}

\noindent{\bf Abstract}

The Wigner representation of parametric down conversion in the Heisenberg picture is applied to the study of the Rome teleportation experiment. We investigate the physical meaning of the zeropoint inputs at the different areas of the experimental setup. In particular, we establish a quantitative relationship between the zeropoint sets of modes that are needed for the preparation of the quantum state to be teleported, the idle channels inside the one-photon polarization-momentum Bell-state analyser, and the possibility of performing teleportation of a polarization state whith certainty.

\vspace{1cm}
\newpage
\section{Introduction}

Since the beginning of the quantum information theory, the possibility of transferring an unknown quantum state by using the fundamental properties of quantum mechanics, has represented one of the most important goals to achieve \cite{In3}. Quantum teleportation is based on two distinctive features of quantum mechanics: entanglement and the projection postulate. These properties, along with the classical communication channel between Alice (the sender) and Bob (the receiver), allows for the possibility of performing teleportation. Nowadays, teleportation constitutes an essential piece for the development of quantum computing and quantum communication \cite{A1,tr4,tr5,Graham}.
Similar to quantum teleportation is remote state preparation, in which Alice has complete classical knowledge of the state she wants to transmit. This distinction implies that remote state preparation reveals a non trivial trade-off between entanglement and classical communication \cite{A,B,C}. 

In the last few decades, the process of parametric down conversion (PDC) has been used for numerous experiments on conceptual problems of quantum mechanics \cite{B1,B2} and optical quantum communication and information processing \cite{B3,B4,B5}. Concretely, the first experimental implementations of quantum teleportation were performed during the nineties by using down converted photons \cite{PDC1,14}. Other optical implementations of quantum teleportation include \cite{15,16,17,nueva}.

In the standard theoretical approach of teleportation, Alice has a two-state system in an unknown quantum state to be teleported (qubit 1). In order to carry out teleportation, Alice and Bob share a bipartite system (particles $2$ and $3$) in an entangled state. A Bell-state measurement (BSM) on the joint system consisting of the unknown state and particle 2 at Alice's station gives Alice a piece of information which, transmitted to Bob, allows him to perform one of four possible unitary transformations on particle $3$ in order to reconstruct the unknown input state \cite{In3}. The experimental realization was carried out
in Innsbruck by using two independent pairs of down converted photons,
and one of the four
photons as a trigger, in order to generate the state to be teleported \cite{PDC1}. The
impossibility to perform a complete Bell-state analysis of two photons,
using entanglement only in one degree of freedom and linear optics \cite{B4}, implies that this scheme for
teleportation cannot be accomplished with 100\% success, even in the ideal
situation of perfect detectors.

In contrast, in the Rome teleportation experiment only two entangled photons are used, and the qubit to be teleported is encoded in one of two degrees of freedom (polarization or momentum) of Alice's photon \cite{14}. The other photon of the down converted pair is sent to Bob. The advantage of this teleportation scheme is that a complete BSM of one-photon polarization-momentum Bell-states is possible. Nevertheless, the input state cannot be supplied by an external system, and this presents some limitations, such as the inability to teleport entangled or mixed states \cite{A1}. The result obtained by Alice is communicated to Bob, who uses this information to apply one out of four unitary transformations, in order to reproduce the original state. This experiment was proposed by Popescu \cite{13} and performed in Rome \cite{14} by using two-photon momentum entanglement, and polarization coding for preparing the qubit to be teleported.

The Innsbruck and Rome experiments omitted the final stage of teleportation, the unitary transformations applied by Bob after the classical communication in order to reconstruct the unknown state. This  was accomplished using nuclear magnetic resonance \cite{Niel}, and later the first long-distance optical quantum teleportation experiment with active feed-forward in real time was accomplished \cite{Ma}.

The Wigner representation of quantum optics in the Heisenberg picture
(WRHP) has been applied in recent years to the study of experiments
on quantum communication using photons generated via parametric
down conversion. The WRHP formalism of PDC resembles nonlinear classical
optics, by taking into consideration the vacuum inputs at the nonlinear
crystal and the different linear optical devices placed between the source of
down converted photons and the detectors. The zeropoint field (ZPF) appears as a
stochastic field that couples with the laser beam into the crystal, giving rise
to the down converted beams \cite{PDC7,pdc4}. In this way, the signals emitted by the crystal
constitute needles of radiation generated by the amplification of vacuum
fluctuations \cite{E1}. These signals propagate according to the classical Maxwell equations, so that the WRHP formalism of PDC emphasizes the wave-like aspects in the generation and propagation of light \cite{PDC8}. The Wigner function of PDC is positive, and corresponds to the Gaussian Wigner distribution of the vacuum state. In contrast, the typical
quantum behaviour of PDC light is related to the way in which vacuum fluctuations
are subtracted at the detectors. The zeropoint intensity appears just
as a threshold for detection, in such a way that signals are detected because
its mean intensity is above the zeropoint background.

The WRHP formalism has been applied to the analysis of the influence
of ZPF inputs in experiments on quantum cryptography \cite{cr}, partial BSM \cite{CD}, hyperentanglement and complete BSM \cite{Hy}, and entanglement swapping \cite{sw}. Two features of the ZPF constitute the common denominator in these works. On the one hand, the quantum information contained in the quantum state of light is stored into the ZPF inputs that are amplified at the source, so that these amplitudes carry the quantum information through the experimental setup. On the other hand, the ZPF inputs at the idle channels inside the analysers constitute an essential source of noise that limits the information that can be extracted in the measurement process. Concretely, in \cite{Hy} it has been demonstrated  that the
number of mutually distinguishable Bell-state classes corresponding to two-photon
entanglement in $n$ degrees of freedom, which are not brought together at the apparatus, $2^n$ \cite{Pisenti}, is equal to the difference between the total number of ZPF sets of modes that are amplified at the source and enter the analyser, $2^{n+1}$, and the number of idle channels inside the analyser, $2^n$.

The paper is organised as follows: in Section \ref{sec1} we shall describe  the Rome teleportation experiment 
by using the WRHP formalism. We shall put the emphasis on the role of the zeropoint entries at the different steps of the scheme proposed by Popescu \cite{ekert}: the generation of two-photon momentum entanglement, the preparation  of the state to be teleported, the complete Bell-state analysis at Alice's station, and the operations performed at Bob's station in order to reconstruct the original state. Also, we shall analyse the verification measurements corresponding to the teleportation of a linearly polarized state and of an elliptically polarized state in the Rome experiment \cite{14}. In Section \ref{sec4} we analyse the relationship between the number of amplified ZPF sets of modes entering Alice's station, the corresponding ones entering the idle channels inside the analyser, and the optimality of the one-photon polarization-momentum BSM. Furthermore, it is emphasized the asymmetry between Alice and Bob with respect to the number of ZPF sets of modes that intervene on each station and its relationship with the necessity of classical information in teleportation. Finally, in Section \ref{sec5} we shall present the main conclusions of the work. In addition, in Appendix \ref{A} we have included  a brief summary of the most important aspects of the WRHP formalism of PDC, which will be used in this paper. In Appendix \ref{B} we describe the basic ideas of the Rome experiment in the standard Hilbert space formalism, in order to compare with the description using the WRHP approach.
\section{The Rome teleportation experiment in the WRHP formalism}\label{sec1}
In this Section we shall analyse the Rome teleportation experiment in the WRHP formalism, by focusing on the role of the  ZPF inputs at the sketch of the experimental setup proposed by  Popescu \cite{13}. Although the experimental implementation \cite{14} was performed by using a slightly different setup to the one proposed in \cite{13}, we have preferred to analyse the original proposal, which does not influence the main results of the paper.
\subsection{The source of two-photon momentum entanglement}
\begin{figure}[H]
      \centering
      \includegraphics[width=12.0cm,clip]{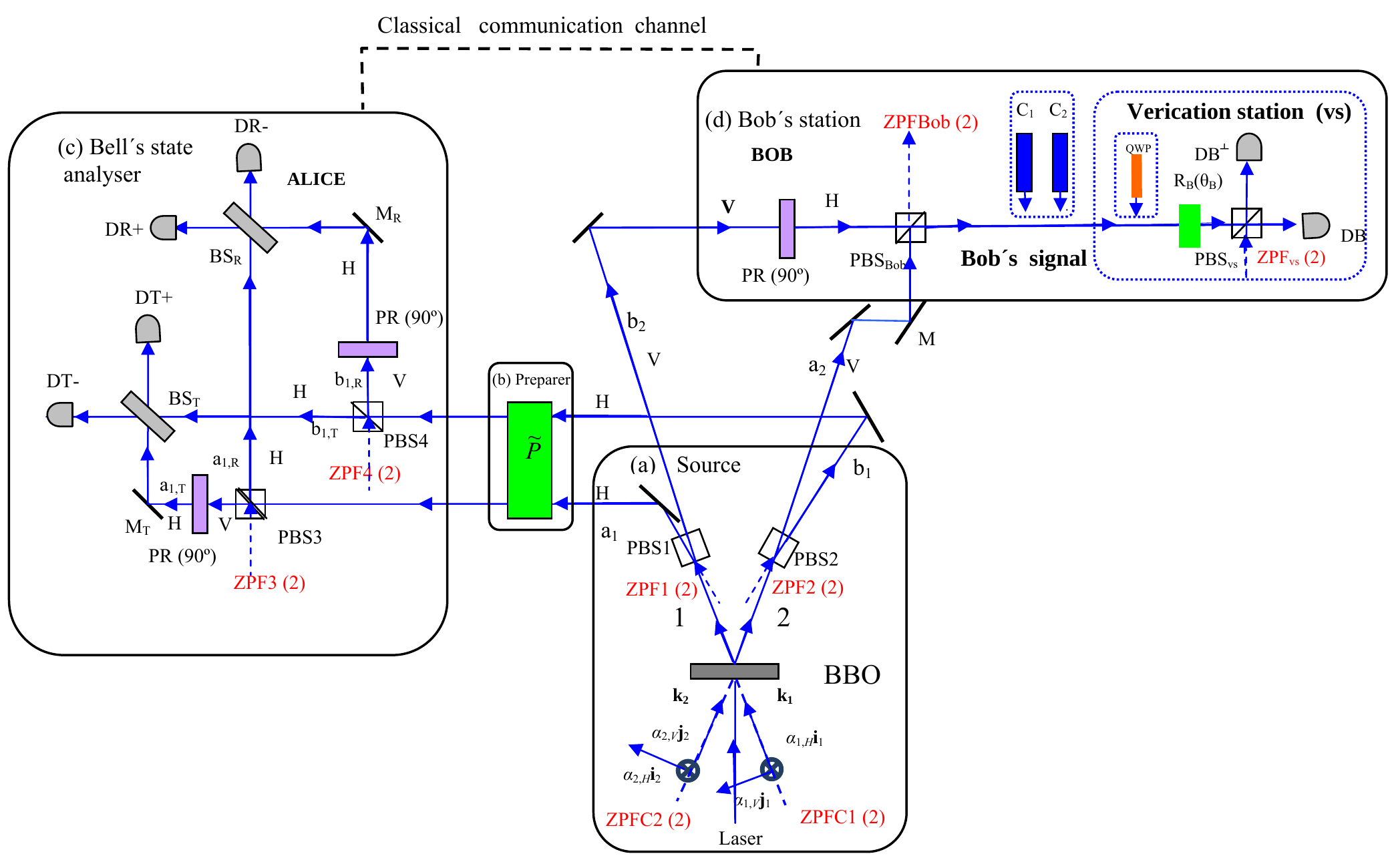}
\caption{\footnotesize{Vacuum inputs corresponding to the Popescu's scheme for teleportation \cite{ekert}. The information concerning the number of ZPF sets of modes is written between brackets. (a) Source: The inputs ZPFC1 and ZPFC2, each of them containing two sets of ZPF modes, are ``activated" and coupled with the laser inside the crystal, so that polarization entanglement is generated. The inputs ZPF1 and ZPF2, entering polarizing beam-splitters PBS1 and PBS2 respectively, introduce four additional sets of ZPF modes, which are necessary  for transferring polarization entanglement to momentum entanglement. (b) Preparer: There are no additional ZPF inputs in the preparation of the state to be teleported. (c) Bell-state analyser: The inputs ZPF3 and ZPF4, entering polarizing beam-splitters PBS3 and PBS4 respectively, introduce four sets of ZPF modes which contribute to the extraction of the information concerning each of the four one-photon polarization-momentum Bell-states. (d) Bob's station: The role of $\rm PBS_{Bob}$ is to transform the cross-correlations between  the field amplitudes at the detectors at Alice's station and the momentum of Bob's photon, to the cross-correlations involving the field amplitudes at Alice's detectors and  the polarization components of Bob's signal. The output $\rm ZPF_{Bob}$ eliminates the excess of noise at Bob's station, and the optical devices $C_{1}$ and $C_{2}$ can be used in order to reproduce the prepared state, after the classical communication between Alice and Bob. In the experimental implementation Bob's signal enters a verification station (vs) constituted by a polarization rotator $\rm R_{B}(\theta_{B})$ and a $\rm PBS_{vs}$, followed by two detectors $\rm D{B}$ and $\rm D{B}^{\bot}$, in the case of the teleportation of a linearly polarized state. In the situation corresponding to the teleportation of an elliptically polarized state a quarte-wave plate QWP must be placed before $\rm R_{B}(\theta_{B})$} \cite{14}.}
\label{Figure1}\end{figure}

The quantum predictions corresponding to the two-photon polarization entangled state, given in Equation (\ref{Hilbert1}), are reproduced in the WRHP approach through the consideration of the following two correlated beams (see Equations (\ref{hyper0}) and (\ref{hyper3})):
\begin{equation} \begin{array}{l} {\bf{F}}_{1}^{(+)} ({\bf{r}}_{C},t)=F_{s}^{(+)} ({\bf{r}}_{C},t;\{\alpha _{{\bf{k}}_{1},H};\alpha ^{*}_{{\bf{k}}_{2}, V }\}){{\bf{i}}_{1}}+F_{p}^{(+)} ({{\bf{r}}_{C}},t;\{\alpha _{{\bf{k}}_{1},V };\alpha^{*} _{{\bf{k}}_{2},H } \}){{\bf{j}}_{1}},\end{array}\label{d1}\end{equation}
\begin{equation} \begin{array}{l} {{\bf{F}}_{2}^{(+)} ({{\bf{r}}_{C}},t)=F_{q}^{(+)} ({{\bf{r}}_{C}},t;\{\alpha _{{\bf{k}}_{2},H};\alpha ^{*}_{{\bf{k}}_{1}, V }\}){\bf{i}}_{2}+F_{r}^{(+)} ({\bf{r}}_{C}},t;\{\alpha _{{\bf{k}}_{2},V };\alpha^{*} _{{\bf{k}}_{1},H }\}){\bf{j}}_{2},\end{array}\label{d2}\end{equation}
which are generated from the coupling inside the non-linear medium between the laser field and the ZPF inputs 
ZPFC1 and ZPFC2 (see Figure \ref{Figure1}). We have considered that the centre of the non-linear source is located at position ${{\bf{r}}_{C}}$. The four sets of ZPF modes that are amplified at the crystal are contained into the beams ${\bf F}_{1}^{(+)}$ and ${\bf F}_{2}^{(+)}$, and each polarization component carries information of two sets of vacuum modes.

Now, beams ``$1$'' and ``$2$'' are passed through polarizing beam-splitters PBS1 and PBS2 respectively, which transmit (reflect) vertical (horizontal) polarization. The idle channels of the PBSs constitute two fundamental inputs of ZPF modes, ZPF1 and ZPF2, in order to ``brake'' the beams ``$1$'' and ``$2$'' for the generation of momentum entanglement from polarization entanglement, which gives rise to the addition of two new momentum degrees of freedom. Each idle channel introduces two sets of vacuum modes, so that the total number of ZPF sets of modes at the source is equal to eight. The two-photon momentum entangled state, given in Equation (\ref{Hilbert2}), is represented in the WRHP formalism by the following four beams: 
\begin{equation}\begin{array}{l}{\bf{F}}_{a_{1} }^{(+)} ({{\bf{r}}_{PBS1}},t)=iF_{s}^{(+)} ({{\bf{r}}_{PBS1}},t){{\bf{i}}_{a_{1}}}+[{\bf{F}}_{ZPF1}^{(+)} ({{\bf{r}}_{PBS1}},t)\cdot{\bf{j}}_{a_{1}}]{\bf{j}}_{a_{1}},\label{5}\end{array}\end{equation}
\begin{equation}\begin{array}{l}{\bf{F}}_{b_{2} }^{(+)} ({{\bf{r}}_{PBS1}},t)=F_{p}^{(+)} ({{\bf{r}}_{PBS1}},t){{\bf{j}}_{b_{2}}}+i[{\bf{F}}_{ZPF1}^{(+)} ({  {\bf{r}}_{PBS1}},t)\cdot{\bf{i}}_{b_{2}}]{\bf{i}}_{b_{2}},\label{1}\end{array}\end{equation}
\begin{equation}\begin{array}{l}{\bf{F}}_{b_{1} }^{(+)} ({{\bf{r}}_{PBS2}},t)=iF_{q}^{(+)} ({{\bf{r}}_{PBS2}},t){{\bf{i}}_{b_{1}}}+[{\bf{F}}_{ZPF2}^{(+)} ({{\bf{r}}_{PBS2}},t)\cdot{\bf{j}}_{b_{1}}]{\bf{j}}_{b_{1}},\label{a}\end{array}\end{equation}
\begin{equation}\begin{array}{l}{\bf{F}}_{a_{2} }^{(+)} ({{\bf{r}}_{PBS2}},t)=F_{r}^{(+)} ({{\bf{r}}_{PBS2}},t){{\bf{j}}_{a_{2}}}+i[{\bf{F}}_{ZPF2}^{(+)} ({{\bf{r}}_{PBS2}},t)\cdot{\bf{i}}_{a_{2}}]{\bf{i}}_{a_{2}},\label{2}\end{array}\end{equation}
with ${{\bf{r}}_{PBS1}}$ (${{\bf{r}}_{PBS2}}$) being the position of PBS1 (PBS2). The beam ${\bf{F}}_{a_{1} }^{(+)}$ (${\bf{F}}_{b_{2} }^{(+)}$) is only correlated to ${\bf{F}}_{a_{2} }^{(+)}$ (${\bf{F}}_{b_{1} }^{(+)}$). This can be easily seen by taking into consideration that $F^{(+)}_{s}$ ($F^{(+)}_{p}$) is only correlated to $F^{(+)}_{r}$ ($F^{(+)}_{q}$), and that the zeropoint inputs ${\bf{F}}^{(+)}_{ZPF1}$ and ${\bf{F}}^{(+)}_{ZPF2}$ are uncorrelated with the signals emitted by the crystal, and with each other. On the other hand, it can be easily seen  the polarization is well-defined at the beams  ${\bf{F}}_{a_{1} }^{(+)}$ and ${\bf{F}}_{b_{1} }^{(+)}$ (horizontal), and also at the beams ${\bf{F}}_{a_{2} }^{(+)}$ and ${\bf{F}}_{b_{2} }^{(+)}$ (vertical).

Beams ${\bf{F}}_{a_{1} }^{(+)}$ and ${\bf{F}}_{b_{1} }^{(+)}$ are directed to the Preparer and beams ${\bf{F}}_{a_{2} }^{(+)}$ and ${\bf{F}}_{b_{2} }^{(+)}$ are sent to Bob. The information transmitted  by the quantum-commu\-nication channel is  carried by the four sets of ZPF amplitudes entering the crystal, along with the four sets of ZPF amplitudes corresponding to the idle channels of PBS1 and PBS2. 

\subsection{The Preparer}
The optical implementation of the superposition given in Equation (\ref{Hilbert3}) is accomphished through the identical action of a set of linear devices placed at each of the beams ${\bf{F}}_{a_{1}}^{(+)}$ and ${\bf{F}}_{b_{1}}^{(+)}$, whose action on the Wigner field amplitudes is treated in the same way as for classical optics. In this case, the Preparer does not introduce additional zeropoint inputs given that the number of degrees of freedom continue to be the same after the preparation. 
To simplify notation we shall, for the time being, discard the dependence on position and time of the field amplitudes but we shall draw the reader's attention, wherever necessary, to any reintroduction of the spacetime variables. Also, from now on we shall use matrix notation for representing vectors.

Let $\hat{P}$ be the matrix representing  the global action of the apparatuses. The outgoing beams ${\bf{F'}}_{a_{1}}^{(+)}$ and ${\bf{F'}}_{b_{1}}^{(+)}$ are:
\begin{equation}\begin{array}{l}{\bf{F'}}_{a_{1}}^{(+)}=\hat{P}{\bf{F}}_{a_{1}}^{(+)}=\left(\begin{array}{cc} A & B \\ C & D \end{array}\right)\left(\begin{array}{c} iF_{s}^{(+)} \\ {F_{ZPF1, V}^{(+)}}\end{array}\right)\\\
=\left(\begin{array}{c} iAF_{s}^{(+)}+B{F_{ZPF1, V}^{(+)}} \\ iCF_{s}^{(+)}+D{F_{ZPF1, V}^{(+)}}\end{array}\right), \label{a1}\end{array}\end{equation}
\begin{equation}\begin{array}{l}{\bf{F'}}_{b_{1}}^{(+)}=\hat{P}{\bf{F}}_{b_{1}}^{(+)}=\left(\begin{array}{cc} A & B \\ C & D \end{array}\right)\left(\begin{array}{c} iF_{q}^{(+)} \\ {F_{ZPF2, V}^{(+)}} \end{array}\right)\\
=\left(\begin{array}{c} iAF_{q}^{(+)}+B{F_{ZPF2, V}^{(+)}}\\ iCF_{q}^{(+)}+D{F_{ZPF2, V}^{(+)}}\end{array}\right),
\label{a2}\end{array}\end{equation}
where ${F_{ZPF1, V}^{(+)}}={{\bf{F}}_{ZPF1}^{(+)}}\cdot {\bf{j}}_{a_{1}}$ and ${F_{ZPF2, V}^{(+)}}={{\bf{F}}_{ZPF2}^{(+)}}\cdot {\bf{j}}_{b_{1}}$. Since $\hat{P}$ represents a unitary transformation, the following relationships must be fulfilled:
\begin{equation}\begin{array}{l}
|{A}|^{2}+|{B}|^{2}=|{C}|^{2}+|{D}|^{2}=1,\label{ya41}\end{array}\end{equation}
  \begin{equation}\begin{array}{l} {A}{C}^{*}+ {B}{D}^{*}=0.\label{ya51}\end{array}\end{equation}

The quantum predictions corresponding to the two-photon state after the preparation (see Equation (\ref{Hilbert4})) are reproduced, in the WRHP formalism, via the beams (\ref{1}), (\ref{2}), (\ref{a1}) and (\ref{a2}). The autocorrelation properties concerning the polarization components of the beams ${\bf{F'}}_{a_{1}}^{(+)}$ and ${\bf{F'}}_{b_{1}}^{(+)}$, at a given point ${\bf{r}}$ and time t, are directly related to the single detection probabilities associated to a given polarization component. By using Equation (\ref{new}), we have:
\begin{equation}
\langle F'^{(+)}_{x_{1}, H}F'^{(-)}_{x_{1}, H}\rangle-\langle F'^{(+)}_{ZPF; x_{1}, H}F'^{(-)}_{ZPF; x_{1}, H}\rangle =|A|^{2}\frac{g^{2}|V|^{2}}{2}\mu_{}(0), 
\label{ya1}
\end{equation}
\begin{equation}
\langle F'^{(+)}_{x_{1}, V}F'^{(-)}_{x_{1}, V}\rangle-\langle F'^{(+)}_{ZPF; x_{1}, V}F'^{(-)}_{ZPF; x_{1}, V}\rangle =|C|^{2}\frac{g^{2}|V|^{2}}{2}\mu_{}(0), 
\label{ya2}
\end{equation}
where $x=\{a, b\}$, and $F'^{(+)}_{ZPF; x_{1}, H}$ ($F'^{(+)}_{ZPF; x_{1}, V}$) represents the contribution of the zeropoint field to the horizontal (vertical) polarization component of  the beam $F'^{(+)}_{x_{1}, H}$ ($F'^{(+)}_{x_{1}, V}$).

Now, we shall compute the non-null cross-correlations between the polarization components of ${\bf{F'}}_{a_{1} }^{(+)}$ and ${\bf{F}}_{a_{2} }^{(+)}$, and also those concerning the beams ${\bf{F'}}_{b_{1} }^{(+)}$ and ${\bf{F}}_{b_{2} }^{(+)}$. We have: 
\begin{equation}\begin{array}{l}\langle {{F'}}_{a_{1},H} ^{(+)}{{F}}_{a_{2},V} ^{(+)}\rangle=iA\langle F_{s}^{(+)}F_{r}^{(+)}\rangle,\label{x}\end{array}\end{equation}
\begin{equation}\begin{array}{l}\langle {{F'}}_{b_{1},V} ^{(+)}{{F}}_{b_{2},V} ^{(+)}\rangle=iC\langle F_{q}^{(+)}F_{p}^{(+)}\rangle, \label{x3}\end{array}\end{equation}
\begin{equation}\begin{array}{l}\langle {{F'}}_{a_{1},V} ^{(+)}{{F}}_{a_{2},V} ^{(+)}\rangle=iC\langle F_{s}^{(+)}F_{r}^{(+)}\rangle,\label{x1}\end{array}\end{equation}
\begin{equation}\begin{array}{l}\langle{{F'}}_{b_{1},H} ^{(+)}{{F}}_{b_{2},V} ^{(+)}\rangle=iA\langle F_{q}^{(+)}F_{p}^{(+)}\rangle.\label{x2}\end{array}\end{equation}
After the preparation, the total number of cross-correlations is duplicated, from two to four, and this is due to the zeropoint amplitudes ${{F}_{ZPF1, V}^{(+)}}$ and ${{F}_{ZPF2, V}^{(+)}}$, in which the Preparer can store quantum information (see Equations (\ref{a1}) and (\ref{a2})). At this point, the total information concernig the preparation is shared by the autocorrelations (see Equations (\ref{ya1}) and (\ref{ya2})), and the cross-correlations given in Equations (\ref{x}) to (\ref{x2}).
\subsection{The one-photon polarization-momentum Bell-sta\-te analyser}
The analyser at Alice's station includes two PBSs, PBS3 and PBS4, which operate in opposite ways. PBS3 (PBS4) is placed in the path of beam $a_{1}$ ($b_{1}$), and transmits the vertical (horizontal)
polarization. A polarization rotator of $90^{\circ}$ is located at each of the outgoing channels $a_{1,T}$ and $b_{1,R}$. After some easy algebra, we obtain the following expression for the field amplitudes after considering the action of the PBSs, the polarization rotators, and the mirrors $\rm M_{T}$ and $\rm M_{R}$ (see Figure \ref{Figure1}):
\begin{equation}\begin{array}{l}{\bf{F}}_{a_{1}, T}^{(+)}=\left(\begin{array}{c} CF_{s}^{(+)}-iD{{F}_{ZPF1, V}^{(+)}}\\ -{{F}_{ZPF3, H}^{(+)}}\end{array}\right),
\end{array}\end{equation}
\begin{equation}\begin{array}{l}{\bf{F}}_{b_{1}, T}^{(+)}=\left(\begin{array}{c} iAF_{q}^{(+)}+B {F}_{ZPF2, V}^{(+)}\\ i{F}_{ZPF4, V}^{(+)}\end{array}\right),
\end{array}\end{equation}
\begin{equation}\begin{array}{l}{\bf{F}}_{a_{1}, R}^{(+)}=\left(\begin{array}{c} -AF_{s}^{(+)}+iB{{F}_{ZPF1, V}^{(+)}}\\ {F}_{ZPF3, V}^{(+)}\end{array}\right),
\end{array}\end{equation}
\begin{equation}\begin{array}{l}{\bf{F}}_{b_{1}, R}^{(+)}=\left(\begin{array}{c} iCF_{q}^{(+)}+D{{F}_{ZPF2, V}^{(+)}}\\ i{F}_{ZPF4, H}^{(+)}\end{array}\right),
\end{array}\end{equation}
where ${{\bf{F}}_{ZPF3}^{(+)}}$ and ${{\bf{F}}_{ZPF4}^{(+)}}$ represent the zeropoint beams entering the polarizing beam-splitters PBS3 and PBS4, respectively. Now, beams ${{\bf{F}}_{a_{1},R}^{(+)}}$ and ${{\bf{F}}_{b_{1},R}^{(+)}}$ (${{\bf{F}}_{a_{1},T}^{(+)}}$ and ${{\bf{F}}_{b_{1},T}^{(+)}}$) incide onto a balanced non polarizing beam-splitter $\rm BS_R$ ($\rm BS_T$). Because there is no idle channel at the $\rm BS_{s}$, the beams are transformed without the activation of additional zeropoint modes.
The field amplitudes at the detectors $\rm DR{-}$, $\rm DR{+}$, $\rm DT{-}$ and $\rm DT{+}$ are:
\begin{equation}\begin{array}{l}{\bf{F}}_{DT{\pm}}^{(+)} =\frac{1}{\sqrt{2}}[i^{n_{\pm}}{\bf{F}}_{b_{1}, T}^{(+)}+i^{n_{\mp}}{\bf{F}}_{a_{1}, T}^{(+)}]\\
=\frac{1}{\sqrt{2}}\left(\begin{array}{cc}i^{n_{\pm}+1} AF_{q}^{(+)}+i^{n_{\mp}}CF_{s}^{(+)}+i^{n_{\pm}}BF_{ZPF2, V}-i^{n_{\mp}+1}DF_{ZPF1, V}\\ i^{n_{\pm}+1}F_{ZPF4, V}-i^{n_{\mp}}F_{ZPF3, H} \end{array}\right), \label{R}\end{array}\end{equation} 
\begin{equation}\begin{array}{l}{\bf{F}}_{DR{\pm}}^{(+)} =\frac{1}{\sqrt{2}}[i^{n_{\pm}}{\bf{F}}_{a_{1}, R}^{(+)}+i^{n_{\mp}}{\bf{F}}_{b_{1}, R}^{(+)}]\\
=\frac{1}{\sqrt{2}}\left(\begin{array}{cc} -i^{n_{\pm}}AF_{s}^{(+)}+i^{n_{\mp}+1}CF_{q}^{(+)}+i^{n_{\pm}+1}BF_{ZPF1, V}+i^{n_{\mp}}DF_{ZPF2, V}\\ i^{n_{\pm}}F_{ZPF3, V}+i^{n_{\mp}+1}F_{ZPF4, H} \end{array}\right), \label{T}\end{array}\end{equation}
where $n_{+}=1$ and $n_{-}=0$. 

Let us calculate the single detection probabilities corresponding to the detectors $\rm DX\pm$ $(X=T, R)$. Using the autocorrelation properties of the light field given in Equation (\ref{new}), the expression for the single detection probability (Equation (\ref{probsimple})), and the field amplitudes at the detectors given in Equations (\ref{R}) and (\ref{T}), it can be easily demonstrated that the single detection probabilities are identical, and independent of the parameters $A$, $B$, $C$, and $D$:
\begin{equation}\begin{array}{l}P_{DX\pm}=k_{DX\pm}\frac{g^{2} |V|^{2}}{4}\mu(0),\end{array}\end{equation}
where $k_{DX\pm}$ is a constant related to the efficiency of the detector $\rm DX\pm$. This contrasts to the autocorrelation properties of the beams ${\bf{F}}_{a_{1}}^{(+)}$ and ${\bf{F}}_{b_{1}}^{(+)}$ given in Equations (\ref{ya1}) and (\ref{ya2}), which depend on the parameters $|A|$ and $|C|$. 

Let us now study the cross-correlation properties of the light field. Taking into account that the vacuum amplitudes ${\bf{F}}_{ZPFN}^{(+)}$ ($N=1, 2, 3, 4$) are uncorrelated with the signals and with each other, the only non-null cross-correlations are those concerning the horizontal polarization component of $\rm DX\pm$ $(X=T, R)$ and the vertical polarization component of the beams ${\bf{F}}_{b_{2}}^{(+)}$ and ${\bf{F}}_{a_{2}}^{(+)}$. The following eight cross-correlations contain the intrinsic nature of teleportation. By using Equations (\ref{R}), (\ref{T}), (\ref{1}) and (\ref{2}), we have:
\begin{equation}\begin{array}{l}\langle {{F}}_{DR\pm, H}^{(+)} {{F}}_{a_{2}, V }^{(+)}\rangle=-\frac{i^{n_{\pm}}}{\sqrt{2}}A\langle {{F}}_{s}^{(+)}{{F}}_{r} ^{(+)}\rangle,\label{v}\end{array}\end{equation}
\begin{equation}\begin{array}{l}\langle {{F}}_{DR\pm, H }^{(+)} {{F}}_{b_{2}, V }^{(+)}\rangle=\frac{i^{n_{\mp}+1}}{\sqrt{2}}C\langle {{F}}_{q}^{(+)}{{F}}_{p} ^{(+)}\rangle,\end{array}\end{equation}
\begin{equation}\begin{array}{l}\langle {{F}}_{DT\pm, H }^{(+)}{{F}}_{a_{2}, V }^{(+)}\rangle=\frac{i^{n_{\mp}}}{\sqrt{2}}C\langle {{F}}_{s}^{(+)}{{F}}_{r} ^{(+)}\rangle,\end{array}\end{equation}
\begin{equation}\begin{array}{l}\langle {{F}}_{DT\pm, H }^{(+)}{{F}}_{b_{2}, V }^{(+)}\rangle=\frac{i^{n_{\pm}+1}}{\sqrt{2}}A\langle {{F}}_{q}^{(+)}{{F}}_{p} ^{(+)}\rangle.\label{u}\end{array}\end{equation}
The distinguishing feature of the above Equations is that the momentum at Alice's detectors is correlated to the momentum of Bob's photon.
For a given detector $\rm DX\pm$ ($X=\{T, R\}$), there are two cross-correlations, $\langle {{F}}_{DX\pm, H }^{(+)}{{F}}_{a_{2}, V} ^{(+)}\rangle$ and $\langle {{F}}_{DX\pm, H }^{(+)}{{F}}_{b_{2}, V} ^{(+)}\rangle$, each of which is proportional to $A$ or $C$. Equations (\ref{v}) to (\ref{u}) are similar to the cross-correlations of the light field before Alice's station, which are given in Equations (\ref{x}) to (\ref{x2}). Nevertheless, the information contained in the electromagnetic field before Alice's station is divided into the autocorrelations (Equations (\ref{ya1}) and (\ref{ya2})) and the cross-correlations (Equations (\ref{x}) to (\ref{x2})). In contrast, once we have considered the action of the different optical devices placed at Alice's station, we can state that the total information about the preparation is contained into the eight cross-correlations given Equations (\ref{v}) to (\ref{u}).

\subsection{Bob's station}
Let us analyse Bob's actions in the WRHP approach. The next step consists in transferring the correlation properties given in Equations (\ref{v}) to (\ref{u}) from the momentum degree of freedom into the polarization. He first rotates $90^{\circ°}$ the beam ${\bf{F}}_{b_{2}}^{(+)}$ (see Equation (\ref{1})):
\begin{equation}\begin{array}{l}{\bf{F'}}_{b_{2} }^{(+)} =\left(\begin{array}{c} -F_{p}^{(+)} \\ i{F_{ZPF1, H}^{(+)}} \end{array}\right).\label{k}\end{array}\end{equation}
Now, ${\bf{F'}}_{b_{2}}^{(+)}$ and ${\bf{F}}_{a_{2}}^{(+)}$ (previously reflected at mirror M) are combined at $\rm PBS_{Bob}$ that transmits (reflects) the horizontal (vertical) polarization component (see Figure \ref{Figure1}). By using Equations (\ref{k}) and (\ref{2}), we obtain the following field amplitude for the signal emitted by $\rm PBS_{Bob}$: 
\begin{equation}\begin{array}{l}{\bf{F}}_{Bob }^{(+)}\equiv{\bf{F}}_{Bob, signal }^{(+)}=-\left(\begin{array}{c} F_{p}^{(+)} \\ F_{r}^{(+)} \end{array}\right),\label{40}\end{array}\end{equation}
while, at the other channel of $\rm PBS_{Bob}$, the outgoing beam includes the horizontal polarization components of the ZPF inputs ${\bf{F}}_{ZPF1}^{(+)}$ and  ${\bf{F}}_{ZPF2}^{(+)}$:
\begin{equation}\begin{array}{l}{\bf{F}}_{Bob, noise }^{(+)} =-\left(\begin{array}{c} {F}_{ZPF2,H}^{(+)} \\ {{{F}}_{ZPF1, H}^{(+)}} \end{array}\right).\label{noise}\end{array}\end{equation}
The field amplitudes at Alice's detectors (Equations (\ref{R}) and (\ref{T})) and the beam given in Equation (\ref{40}) correspond to the WRHP description of the state given in Equation (\ref{pura}). The analysis of the correlation properties of the field will allow for a wave-like description of the Rome experiment on teleportation.

 By using Equations (\ref{40}) and (\ref{new}), it can be easily seen that the autocorrelation properties, at a given point ${\bf{r}}$ and time t, concerning the polarization components of ${\bf{F}}_{Bob }^{(+)}$ are equal. We have:
 \begin{equation}\begin{array}{l}
\langle F^{(+)}_{Bob, H}F^{(-)}_{Bob, H}\rangle-\langle F^{(+)}_{ZPF; Bob, H}F^{(-)}_{ZPF; Bob, H}\rangle \\\\
=\langle F^{(+)}_{Bob, V}F^{(-)}_{Bob, V}\rangle-\langle F^{(+)}_{ZPF; Bob, V}F^{(-)}_{ZPF; Bob, V}\rangle =\frac{g^{2}|V|^{2}}{2}\mu_{}(0). 
\label{joycor}\end{array}\end{equation}
These autocorrelations differ from the corresponding ones to ${\bf{F'}}_{a_{1} }^{(+)}$ and ${\bf{F'}}_{b_{1}}^{(+)}$ (see Equations (\ref{ya1}) and (\ref{ya2})), which depend on the values of $|A|$ and $|C|$. Hence, the autocorrelation properties corresponding to ${\bf{F}}_{Bob }^{(+)}$ do not contain any information about the preparation given in Equations (\ref{a1}) and (\ref{a2}). 

The cross-correlation functions concerning the polarization components of ${\bf{F}}_{Bob }^{(+)}$ and the horizontal polarization component of ${\bf{F}}_{DX\pm }^{(+)}$, can be obtained from Equations (\ref{v}) to (\ref{u}), by making the exchange ${{F}}_{b_{2},V }^{(+)}\rightarrow {{F}}_{Bob,H }^{(+)}$, ${{F}}_{a_{2},V }^{(+)}\rightarrow {{F}}_{Bob,V }^{(+)}$, ${{F}}_{p }^{(+)}\rightarrow {{-F}}_{p }^{(+)}$ and ${{F}}_{r }^{(+)}\rightarrow {-{F}}_{r }^{(+)}$. We have:
\begin{equation}\begin{array}{l}\langle {{F}}_{DR\pm, H }^{(+)} {{F}}_{Bob, V} ^{(+)}\rangle=\frac{i^{n_{\pm}}}{\sqrt{2}}A\langle {{F}}_{s}^{(+)}{{F}}_{r} ^{(+)}\rangle,\label{ok1}\end{array}\end{equation}
\begin{equation}\begin{array}{l}\langle {{F}}_{DR\pm, H }^{(+)} {{F}}_{Bob, H} ^{(+)}\rangle=-\frac{i^{n_{\mp}+1}}{\sqrt{2}}C\langle {{F}}_{q}^{(+)}{{F}}_{p} ^{(+)}\rangle,\label{ok} \end{array}\end{equation}
\begin{equation}\begin{array}{l}\langle {{F}}_{DT\pm, H }^{(+)}{{F}}_{Bob, V} ^{(+)}\rangle=-\frac{i^{n_{\mp}}}{\sqrt{2}}C\langle {{F}}_{s}^{(+)}{{F}}_{r} ^{(+)}\rangle.\label{45}\end{array}\end{equation}
\begin{equation}\begin{array}{l}\langle {{F}}_{DT\pm, H }^{(+)}{{F}}_{Bob, H} ^{(+)}\rangle=-\frac{i^{n_{\pm}+1}}{\sqrt{2}}A\langle {{F}}_{q}^{(+)}{{F}}_{p} ^{(+)}\rangle,\label{44}\end{array}\end{equation}

A quick look at Equations (\ref{45}) and (\ref{44}) shows that, in the case of the detector $\rm DT{-}$, the cross-correlation properties are the same to the ones corresponding to Equations (\ref{x}) to (\ref{x2}), except for a constant:
\begin{equation}\begin{array}{l}\langle {{F}}_{DT-, H }^{(+)}{{F}}_{Bob, H}^{(+)}\rangle=-i\frac{\langle {F}_{q}^{(+)}{F}_{p}^{(+)}\rangle}{\sqrt{2}}A,\end{array}\end{equation}
\begin{equation}\begin{array}{l}\langle {{F}}_{DT-, H }^{(+)}{{F}}_{Bob, V} ^{(+)}\rangle=-i\frac{\langle F_{s}^{(+)}{F}_{r}^{(+)}\rangle}{\sqrt{2}}C.\end{array}\end{equation}
Nevertheless, in the case of $\rm DT{+}$, $\rm DR{+}$ and $\rm DR{-}$, the two correlations differ from those given in Equations (\ref{x}) to (\ref{x2}). In the standard Hilbert space formalism, once Alice informs Bob about which detector to click, he applies a unitary transformation in order to resurrect the unknown input state. He uses  two optical devices $C_{1}$ and $C_{2}$, whose action on the Wigner field amplitudes is treated in the same way as for classical optics. For instance, we shall consider the application of a linear transformation $\tilde{M}$ to the beam given in Equation (\ref{40}). We have:
 \begin{equation}\begin{array}{l}{\bf{F'}}_{Bob}^{(+)}=\tilde{M}{\bf{F}}_{Bob} ^{(+)}=\left(\begin{array}{c} \tilde{A} \, \, \, \, \, \tilde{B} \\ \tilde{C }\, \, \, \, \, \tilde{D} \end{array}\right)\left(\begin{array}{c} {-F_{p}^{(+)}}  \\ {-F_{r}^{(+)}} \end{array}\right)=\left(\begin{array}{c} {F'}_{Bob, H}^{(+)}  \\ {F'}_{Bob, V}^{(+)} \end{array}\right),\label{didac}\end{array}\end{equation}
where 
\begin{equation}\begin{array}{l}{F'}_{Bob, H}^{(+)}=-(\tilde{A}{F}_{p}^{(+)}+\tilde{B}{F}_{r}^{(+)}), \label{49}\end{array}\end{equation}
\begin{equation}\begin{array}{l}{F'}_{Bob, V}^{(+)}=-(\tilde{C}{F}_{p}^{(+)}+\tilde{D}{F}_{r}^{(+)}). \label{50}\end{array}\end{equation}
Because of $\tilde{M}$ represents a unitary transformation, similar relationships to the ones provided by Equations (\ref{ya41}) and (\ref{ya51}) must be fulfilled, by making the exchange ($A, B, C, D$)$\rightarrow$($\tilde{A}, \tilde{B}, \tilde{C}, \tilde{D}$).

At this point, we recall that quantum state transmission is not accomplished faster than light, because Bob must wait for Alice's measurement result to arrive before he can recover the quantum state. In the WRHP formalism, this implies that the cross-correlations must involve the field amplitude at Alice's detector $\rm DX\pm$ at time $t_{A}$ (see Equations (\ref{R}) and (\ref{T})), and the transformation of the field amplitudes at Bob's station (see Equations (\ref{didac}) to (\ref{50})), at time $t_{B}=t_{A}+T_{cc}$, $T_{cc}$ being the time interval corresponding to the classical communication between Alice and Bob. That is, at this point the following spacetime labels will be associated to the field amplitudes:
 \begin{equation}\begin{array}{l} {{F}}_{DX\pm, H }^{(+)}\equiv{{F}}_{DX\pm, H }^{(+)}({\bf r}_{DX\pm},t_{A}){\, \, \, \,\, \,};{\, \, \, \,\, \,} X=\{T, R\},\label{time}\end{array}\end{equation}
\begin{equation}\begin{array}{l}{{F'}}_{Bob,z} ^{(+)}={{F'}}_{Bob,z}^{(+)}({\bf{r}}_{B}, t_{B}=t_{A}+T_{cc} ){\, \, \, \,\, \,};{\, \, \, \,\, \,}z=\{H, V\}.\label{time1}\end{array}\end{equation}

 The parameters $\tilde{A}$, ${\tilde{B}}$, $\tilde{C}$ and $\tilde{D}$ must be adjusted, so that the following cross-correlation properties hold:
 \begin{equation}\begin{array}{l}\langle {{F}}_{DX\pm, H }^{(+)}{{F'}}_{Bob, H} ^{(+)}\rangle=\frac{iA}{\sqrt{2}}C_{DX\pm} \langle {{F}}_{q}^{(+)}{{F}}_{p} ^{(+)}\rangle,\label{51}\end{array}\end{equation}
\begin{equation}\begin{array}{l}\langle {{F}}_{DX\pm, H }^{(+)}{{F'}}_{Bob, V} ^{(+)}\rangle=\frac{iC}{\sqrt{2}}C_{DX\pm} \langle {{F}}_{s}^{(+)}{{F}}_{r} ^{(+)}\rangle,\label{52}\end{array}\end{equation}
where 
$C_{DX\pm}$ is an irrelevant constant associated to the detector $\rm DX\pm$. By using Equations (\ref{R}), (\ref{T}), (\ref{49}) and (\ref{50}), we easily obtain:
\begin{equation}\begin{array}{l}\langle {{F}}_{DT\pm, H }^{(+)} {{F'}}_{Bob, H }^{(+)}\rangle=-\frac{i^{n_{\pm}+1}}{\sqrt{2}}(A\tilde{A}\langle {{F}}_{q}^{(+)}{{F}}_{p} ^{(+)}\rangle\mp C\tilde{B}\langle {{F}}_{s}^{(+)}{{F}}_{r} ^{(+)}\rangle),\label{53}\end{array}\end{equation}
\begin{equation}\begin{array}{l}\langle {{F}}_{DT\pm, H }^{(+)} {{F'}}_{Bob, V }^{(+)}\rangle=-\frac{i^{n_{\pm}+1}}{\sqrt{2}}(A\tilde{C}\langle {{F}}_{q}^{(+)}{{F}}_{p} ^{(+)}\rangle\mp C\tilde{D}\langle {{F}}_{s}^{(+)}{{F}}_{r} ^{(+)}\rangle),\label{54}\end{array}\end{equation}
\begin{equation}\begin{array}{l}\langle {{F}}_{DR\pm, H }^{(+)} {{F'}}_{Bob, H }^{(+)}\rangle=\frac{i^{n_{\pm}}}{\sqrt{2}}(A\tilde{B}\langle {{F}}_{s}^{(+)}{{F}}_{r} ^{(+)}\rangle\mp C\tilde{A}\langle {{F}}_{q}^{(+)}{{F}}_{p} ^{(+)}\rangle),\label{55}\end{array}\end{equation}
\begin{equation}\begin{array}{l}\langle {{F}}_{DR\pm, H }^{(+)} {{F'}}_{Bob, V }^{(+)}\rangle=\frac{i^{n_{\pm}}}{\sqrt{2}}(A\tilde{D}\langle {{F}}_{s}^{(+)}{{F}}_{r} ^{(+)}\rangle\mp C\tilde{C}\langle {{F}}_{q}^{(+)}{{F}}_{p} ^{(+)}\rangle),\label{56}\end{array}\end{equation} 
where, in Equations (\ref{51}) to (\ref{56}), the spacetime labels corresponding to the amplitudes $F_{q}^{(+)}$ and $F_{s}^{(+)}$ ($F_{p}^{(+)}$ and $F_{r}^{(+)}$) are those given in Equation (\ref{time})((\ref{time1})).
Now, we shall obtain the appropriate transformation, depending on the outcome of Alice's measurement. We shall consider the following cases: 
\begin{itemize}
\item \textit{Case I }($\rm DT+$ and $\rm DT-$). By substituting (\ref{51}) and (\ref{52}) (by putting $X \equiv T$), into Equations (\ref{53}) and (\ref{54}) respectively, we easily obtain:
  \begin{equation}\begin{array}{l} -i^{n_{\pm}}(A\tilde{A}\langle {{F}}_{q}^{(+)}{{F}}_{p} ^{(+)}\rangle\mp C\tilde{B}\langle {{F}}_{s}^{(+)}{{F}}_{r} ^{(+)}\rangle)=AC_{DT\pm}\langle {{F}}_{q}^{(+)}{{F}}_{p} ^{(+)}\rangle,\label{57}\end{array}\end{equation}
\begin{equation}\begin{array}{l} -i^{n_{\pm}}(A\tilde{C}\langle {{F}}_{q}^{(+)}{{F}}_{p} ^{(+)}\rangle\mp C\tilde{D}\langle {{F}}_{s}^{(+)}{{F}}_{r} ^{(+)}\rangle)=CC_{DT\pm}\langle {{F}}_{s}^{(+)}{{F}}_{r} ^{(+)}\rangle.\label{58}\end{array}\end{equation}

From Equations (\ref{57}) and (\ref{58}) we deduce that $\tilde{B}=\tilde{C}=0$, $\tilde{D}=(-1)^{n_{\pm}}\tilde{A}$, and $C_{DT\pm}=-i^{n_{\pm}}\tilde{A}$. Taking into consideration that $|\tilde{A}|^{2}=1$, and making the choice $\tilde{A}=1$, we obtain:
\begin{equation}\begin{array}{l}\tilde{M}_{DT-}=\hat{I}=\left(\begin{array}{c} {1} \, \, \, \, \, {0} \\ {0}\, \, \, \, \, {1} \end{array}\right)\,\,\ ; \,\,\ C_{DT-}=-1, \end{array}\label{ya6}\end{equation}
\begin{equation}\begin{array}{l}\tilde{M}_{DT+}=\hat{\phi}(\pi)=\left(\begin{array}{cc} {1} & {0} \\ {0} & {{e^{i\pi}}} \end{array}\right)\,\,\ ; \,\,\ C_{DT+}=-i, \end{array}\label{ya7}\end{equation}
where $\hat{I}$ is the identity matrix and $\hat{\phi}(\pi)$ the matrix representing a wave retarder corresponding to a phase change $\pi$ between the vertical and  horizontal polarization components.

\item \textit{Case II }($\rm DR+$ and $\rm DR-$). By substituting (\ref{51}) and (\ref{52}) (by putting $X \equiv R$), into Equations (\ref{55}) and (\ref{56}) respectively, we have:
   \begin{equation}\begin{array}{l} i^{n_{\pm}}(A\tilde{B}\langle {{F}}_{s}^{(+)}{{F}}_{r} ^{(+)}\rangle\mp C\tilde{A}\langle {{F}}_{q}^{(+)}{{F}}_{p} ^{(+)}\rangle)=iAC_{DR\pm}\langle {{F}}_{q}^{(+)}{{F}}_{p} ^{(+)}\rangle,\label{59}\end{array}\end{equation}
\begin{equation}\begin{array}{l} i^{n_{\pm}}(A\tilde{D}\langle {{F}}_{s}^{(+)}{{F}}_{r} ^{(+)}\rangle\mp C\tilde{C}\langle {{F}}_{q}^{(+)}{{F}}_{p} ^{(+)}\rangle)=iCC_{DR\pm}\langle {{F}}_{s}^{(+)}{{F}}_{r} ^{(+)}\rangle.\label{60}\end{array}\end{equation}
\end{itemize}
From Equations (\ref{59}) and (\ref{60}), and taking into consideration that $\langle {{F}}_{q}^{(+)}{{F}}_{p} ^{(+)}\rangle$ $=\langle {{F}}_{s}^{(+)}{{F}}_{r} ^{(+)}\rangle$ (see Equation \ref{nu}), we have $\tilde{A}=\tilde{D}=0$, $\tilde{C}=(-1)^{{n_{\pm}}}\tilde{B}$, and $C_{DR\pm}=i^{n_{\pm}-1}\tilde{B}$. Because of $|\tilde{B}|^{2}=1$, and making the choice $\tilde{B}=-1$, we have: 
\begin{equation}\begin{array}{l}\tilde{M}_{DR+}=\hat{R}(\frac{\pi}{2})=\left(\begin{array}{cc} {0} & {-1 } \\ {1} & {0} \end{array}\right)\,\,\ ; \,\,\ C_{DR+}=-1, \end{array}\label{ya8}\end{equation}
\begin{equation}\begin{array}{l}\tilde{M}_{DR-}=\hat{\phi}(\pi)\hat{R}(\frac{\pi}{2})=\left(\begin{array}{cc} {0} & {-1 } \\ {-1} & {0} \end{array}\right)\,\,\ ; \,\,\ C_{DR-}=i, \end{array}\label{ya9}\end{equation}
where $\hat{R}(\theta)$ is the matrix corresponding to a polarization rotator of angle $\theta$ with respect to the horizontal direction.

Let $d_{AC}$ be the identical optical path length between the crystal and any of the detectors $\rm D{X\pm}$ placed at Alice's station. And let $d_{BC}$ be the optical path length between the crystal and the position ${\bf r}_{B}$ where Bob activate one of the transformations (\ref{ya6}), (\ref{ya7}), (\ref{ya8}) or (\ref{ya9}). By using Equations (\ref{propagation}) and (\ref{nu}), the following condition must be fulfilled between $T_{cc}$, $d_{AC}$, $d_{BC}$ and the correlation time between the down-converted photons \cite{PDC8}, in order to achieve success in the reconstruction of the cross-correlation properties (see Equations (\ref{51}) and (\ref{52})):
\begin{equation}|T_{cc}+\frac{d_{AC}-d_{BC}}{c}|\leq\tau.\label{coherence}\end{equation}

\subsubsection{The verification station}
In the Rome teleportation experiment \cite{14}, rather than performing the transformations described in Equations (\ref{ya6}), (\ref{ya7}), (\ref{ya8}) and (\ref{ya9}), Bob's signal enters a verification station (vs) in order to check that the correlation properties given in Equations (\ref{ok1})  to (\ref{44}) are fulfilled. We shall describe the WRHP analysis of the situations corresponding to the teleportation  of a linearly polarized state and of an elliptically polarized state.

For the sake of clarity an identical distance separating the crystal from the respective optical devices and detectors will be considered (i.e. $d_{AC}=d_{BC}$), so that the related phase shift in Equation (\ref{propagation}) will be discarded in the calculations. Also, without loss of generality, we shall consider the ideal situation $t_{A}=t_{B}$, so that we can discard the spacetime dependence.
\begin{itemize}
\item Linear polarization.

In this case, $\hat{P}$ is the matrix corresponding to a polarization rotator of angle $\theta$ with respect to the horizontal direction, so that $A=\rm cos\theta$, $B=\rm-sin\theta$, $C=\rm sin\theta$, $D=\rm cos\theta$, where $\theta= 22,5^{o}$. Bob's signal impinges on a polarization rotator $\rm R_{B}(\theta_{B})$, and then it enters a polarizing beam-splitter $\rm PBS_{vs}$ that transmits (reflects) the horizontal (vertical) polarization, followed by two detectors, $\rm D{B}$ and $\rm D{B}^{\bot}$ (in reference \cite{14} the detector $\rm D{B}^{\bot}$ is used only for the alignment of Bob's apparatus). The field amplitudes outgoing  $\rm R_{B}(\theta_{B})$ can be easily obtained by using Equations (\ref{didac}), (\ref{49}) and (\ref{50}), with $\tilde{A}=\rm cos\theta_{B}$, $\tilde{B}=\rm-sin\theta_{B}$, $\tilde{C}=\rm sin\theta_{B}$ and $\tilde{D}=\rm cos\theta_{B}$, and by making the change ${\bf{F'}}_{Bob}^{(+)}\rightarrow{\bf{F}}_{vs}^{(+)}$. We have:
\begin{equation}\begin{array}{l}{\bf{F}}_{vs}^{(+)}\equiv\left(\begin{array}{c} {F}_{vs, H}^{(+)} \\ {F}_{vs, V}^{(+)}\end{array}\right)=\left(\begin{array}{c} {-F_{p}^{(+)}}{\rm cos}\theta_{B}{+F_{r}^{(+)}}{\rm sin}\theta_{B} \\ {-F_{p}^{(+)}}{\rm sin}\theta_{B}{-F_{r}^{(+)}}{\rm cos}\theta_{B}  \end{array}\right).\end{array}\end{equation}

The cross-correlations can be obtained from Equations (\ref{53}) to (\ref{56}) by substituting the matrix elements of $\tilde{P}$ and $\tilde{M}$. We have:

\begin{equation}\begin{array}{l}\langle {{F}}_{DT\pm, H }^{(+)} {{F}}_{vs, H }^{(+)}\rangle=-\frac{i^{n_{\pm}+1}}{\sqrt{2}}gV\nu(0) \rm cos(\theta \mp \theta_{B}),\label{bau}\end{array}\end{equation}
\begin{equation}\begin{array}{l}\langle {{F}}_{DT\pm, H }^{(+)} {{F}}_{vs, V }^{(+)}\rangle=\pm\frac{i^{n_{\pm}+1}}{\sqrt{2}}gV\nu(0)\rm sin(\theta \mp\theta_{B}  ),\label{bau1}\end{array}\end{equation}
\begin{equation}\begin{array}{l}\langle {{F}}_{DR\pm, H }^{(+)} {{F}}_{vs, H }^{(+)}\rangle=\mp\frac{i^{n_{\pm}}}{\sqrt{2}}gV\nu(0) \rm sin(\theta \pm\theta_{B}),\label{bau2}\end{array}\end{equation}
\begin{equation}\begin{array}{l}\langle {{F}}_{DR\pm, H }^{(+)} {{F}}_{vs, V }^{(+)}\rangle=\frac{i^{n_{\pm}}}{\sqrt{2}}gV\nu(0)\rm cos(\theta \pm\theta_{B}),\label{bau3}\end{array}\end{equation} 
where we have taken into account that $\langle {{F}}_{q}^{(+)} {{F}}_{p}^{(+)}\rangle=\langle {{F}}_{s}^{(+)} {{F}}_{r}^{(+)}\rangle=gV\nu(0)$ (see Equation (\ref{nu})).

Now, we shall calculate the joint detection probabilities, by using Equation (\ref{p12}) and by taking into account that ${\bf{F}}_{ZPF_{vs}}^{(+)}$ is uncorrelated with ${\bf{F}}_{vs}^{(+)}$. By defining the constants $k_{DB}$ and $k_{DB^{\bot}}$ related to the effective efficiency of the detection process at the verification station, we obtain:
\begin{equation}
\frac{P_{DT\pm,DB}}{k_{DT\pm}k_{DB}}=\frac{P_{DR\mp,DB^{\bot}}}{k_{DR\mp}k_{DB^{\bot}}}=\frac{g^2|V|^2|\nu(0)|^{2}}{2} \rm cos^{2}(\theta\mp\theta_{B}),
\label{probid}
\end{equation}
\begin{equation}
\frac{P_{DR\pm,DB}}{k_{DR\pm}k_{DB}}=\frac{P_{DT\mp,DB^{\bot}}}{k_{DT\mp}k_{DB^{\bot}}}=\frac{g^2|V|^2|\nu(0)|^{2}}{2} \rm sen^{2}(\theta\pm\theta_{B}).
\label{probid2}
\end{equation}
Equations (\ref{probid}) and (\ref{probid2}) are consistent with the graphs of Fig. 2 of reference \cite{14}.

\item Elliptical polarization.

Now, we shall consider the situation in which the Preparer uses a quarter-wave plate in order to prepare an elliptically polarized state. In this case, $A=\cos^{2}\gamma+i\sin^{2}\gamma$, $B=C=(1-i)\rm sin\gamma cos\gamma$ and $D= \sin^{2}\gamma+i\cos^{2}\gamma$, where $\gamma=20^{o}$ corresponds to the orientation of the quarter-wave plate with respect to the horizontal. The verification  station consists on a quarter-wave plate whose orientation depends on the  detector at Alice's side, i.e. $\gamma_{B}=\gamma_{B}(\rm DX_{\pm}$), and  a polarization rotator $R_{B}(\theta_{B})$ followed by a $\rm PBS_{vs}$ and detectors DB and $\rm DB^{\bot}$. The  electric  field outgoing $R_{B}(\theta_{B})$ can be easily obtained by using Equations (\ref{didac}), (\ref{49}) and (\ref{50}) with
\begin{equation}\begin{array}{c}{\bf{F}}_{vs}^{(+)}=\tilde{M}{\bf{F}}_{Bob} ^{(+)}=\hat{R}_{B}(\theta_{B})\hat{M}_{QWP}(\gamma_{B}){\bf{F}}_{Bob}^{(+)}.\end{array}\label{ya10}\end{equation}
The matrix elements of $\tilde{M}$ are: 
\begin{equation}\begin{array}{c}
\tilde{A}=\cos\theta_{B}(\cos^{2}\gamma_{B}-i\sin^{2}\gamma_{B})-\sin\theta_{B}(1+i)\sin\gamma_{B} \cos\gamma_{B},\end{array}\label{xa}\end{equation}
\begin{equation}\begin{array}{c}\tilde{B}=\cos\theta_{B}(1+i)\sin\gamma_{B} \cos\gamma_{B}-\sin\theta_{B}(\sin^{2}\gamma_{B}-i\cos^{2}\gamma_{B}),\end{array}\label{xb}\end{equation}
\begin{equation}\begin{array}{c}\tilde{C}=\sin\theta_{B}(\cos^{2}\gamma_{B}-i\sin^{2}\gamma_{B})+\cos\theta_{B}(1+i)\sin\gamma_{B} \cos\gamma_{B},\end{array}\label{xc}\end{equation}
\begin{equation}\begin{array}{c}\tilde{D}=\sin\theta_{B}(1+i)\sin\gamma_{B} \cos\gamma_{B}+\cos\theta_{B}(\sin^{2}\gamma_{B}-i\cos^{2}\gamma_{B})
.\end{array}\label{xd}\end{equation}
Taking into account that the values of $\gamma_{B}$ corresponding to detectors $\rm DT\pm$ are $\gamma_{B}(DT\pm)=\mp\gamma$, we obtain the following cross-correlations by substituting expressions (\ref{xa}) to (\ref{xd}) into Equations (\ref{53}) and (\ref{54}):
\begin{equation}\begin{array}{l}\langle {{F}}_{DT\pm, H }^{(+)} {{F}}_{vs, H }^{(+)}\rangle=-\frac{i^{n_{\pm}+1}}{\sqrt{2}}gV\nu(0) \rm cos\theta_{B},\label{ya11}\end{array}\end{equation}
\begin{equation}\begin{array}{l}\langle {{F}}_{DT\pm, H }^{(+)} {{F}}_{vs, V }^{(+)}\rangle=-\frac{i^{n_{\pm}+1}}{\sqrt{2}}gV\nu(0) \rm sin\theta_{B}.\label{ya12}\end{array}\end{equation}
On the other hand, the values  of $\gamma_{B}$ corresponding to detectors $\rm DR\pm$ are $\gamma_{B}(DR\pm)=\pm\gamma+90^{o}$. By substituting expressins (\ref{xa}) to (\ref{xd}) into Equations (\ref{55}) and (\ref{56}), we have:
\begin{equation}\begin{array}{l}\langle {{F}}_{DR\pm, H }^{(+)} {{F}}_{vs, H }^{(+)}\rangle=-\frac{i^{n_{\pm}}}{\sqrt{2}}gV\nu(0) \rm sin\theta_{B},\label{ya13}\end{array}\end{equation}
\begin{equation}\begin{array}{l}\langle {{F}}_{DR\pm, H }^{(+)} {{F}}_{vs, V }^{(+)}\rangle=\frac{i^{n_{\pm}}}{\sqrt{2}}gV\nu(0) \rm cos\theta_{B}.\label{ya14}\end{array}\end{equation}
The joint detection probabilities can be calculated by using Equation (\ref{p12}) and by taking into account that ${\bf{F}}_{ZPFvs}^{(+)}$ is uncorrelated with ${\bf{F}}_{vs}^{(+)}$. We obtain:
 \begin{equation}
\frac{P_{DT\pm,DB}}{k_{DT\pm}k_{DB}}=\frac{P_{DR\pm,DB^{\bot}}}{k_{DR\pm}k_{DB^{\bot}}}=\frac{g^2|V|^2|\nu(0)|^{2}}{2} \rm cos^{2}\theta_{B},
\label{ya15}
\end{equation}
\begin{equation}
\frac{P_{DT\pm,DB^{\bot}}}{k_{DT\pm}k_{DB^{\bot}}}=\frac{P_{DR\pm,DB^{}}}{k_{DR\pm}k_{DB^{}}}=\frac{g^2|V|^2|\nu(0)|^{2}}{2} \rm sen^{2}\theta_{B}.
\label{ya16}
\end{equation}
Equations (\ref{ya15}) and (\ref{ya16}) are consistent with the graphs of Fig. 3 of reference \cite{14}.
\end{itemize}
\section{ZPF and complete BSM in the Ro\-me experiment}\label{sec4}
The role of the ZPF inputs in optical quantum communication is closely
related to the area of the experimental setup where they are acting. In the Rome teleportation experiment, the four sets of ZPF modes entering the crystal are amplified to produce two-photon polarization entanglement (see Equations (\ref{d1}) and (\ref{d2})), and the four additional sets of ZPF modes entering PBSs $1$ and $2$ are activated for transferring polarization entanglement to momentum entanglement (see Equations (\ref{5}) to (\ref{2})). Given that the Preparer does not introduce additional zeropoint modes, the full quantum electrodynamical description before 
the BSM at Alice's station is supported by the eight sets of ZPF modes which are necessary for the generation of two-photon momentum entanglement.

The two beams entering  and leaving the Preparer, and the two beams directed to Bob's station, contain the information concerning the four sets of ZPF modes entering the crystal. Concretely, the cross-correlations given in Equations (\ref{x}) to (\ref{x2}) constitute the quantum-communication channel. Additionally, they share the four sets of ZPF amplitudes that enter PBSs 1 and 2: the
reflected (transmitted) ZPF are directed to Alice (Bob). In this way, the two beams entering Alice's station, ${\bf F'}^{(+)}_{a_1}$ and ${\bf F'}^{(+)}_{b_1}$, and the beams entering Bob's station, ${\bf F}^{(+)}_{a_2}$ and ${\bf F}^{(+)}_{b_2}$, contain information concerning six sets of vacuum modes. 

Let us now analyse the role of the ZPF inputs entering the idle channels of the polarizing beam-splitters PBS3 and PBS4 placed at Alice's station (see Figure \ref{Figure1}). These two inputs of noise introduce 
four sets of ZPF modes which contribute to the extraction of  the information concerning each of the four
one-photon polarization-momentum Bell states. The 
polarization rotators, mirrors and beam-splitters placed at Alice's station do
not introduce additional zeropoint modes, so that the total vacuum contribution inside the analyser is represented by four sets of ZPF modes. The net effect of the vacuum inputs inside the Bell-state analyser is to reduce the upper bound corresponding to the classical information  that can be extracted in the measurement, in relation to the six sets of amplified ZPF modes entering Alice's station. 

In order to quantitatively demonstrate the relationship between the zeropoint inputs at the experimental setup and the optimality of the Bell-state analysis, we shall focus on Equation (36) of reference \cite{Hy}. This Equation relates the maximun number of distinguishable Bell-state classes of two photons, entangled in $n$ dichotomic degrees of freedom, in experiments in which the two photons are not mixed at the device, with the number of ZPF inputs at the source of entanglement and inside the analyser. In the standard Hilbert space approach, given to the fact that the first detection event, which is produced in one of the $2^{n}$ detectors of the left (or right) area, does not give any information about the Bell-state of the two photons, the second detection event can discriminate to $2^n$ sets of Bell states \cite{Pisenti}. From the WRHP approach, this quantity can be obtained by subtracting the total number of idle channels at the analyser, $N_{ic}$, from the total number of ZPF sets of modes that are amplified at the source of hyperentanglement, $N_{ZPF, S}$. By taking into consideration that each idle channel introduces two sets of vacuum modes, we have $N_{ic}=N_{ZPF, A}^{noise}/2$, $N_{ZPF, A}^{noise}$ being the total number of sets of vacuum modes entering the $N_{ic}$ entry points of noise inside the analyser. Thus, Equation (36) of reference \cite{Hy} can be written as:

\begin{equation}
N_{max, class}=N_{ZPF, S}-N_{ic}=N_{ZPF, S}-\frac{N_{ZPF, A}^{noise}}{2}.
\label{conclus1}
\end{equation}

Now, let $N_{ZPF, A}$ be the number of amplified ZPF sets of modes entering the analyser. It comes as an inmediate consequence of lemma II of \cite{Hy}, that $N_{ZPF, A}=N_{ZPF, S}$ in the situations concerning the Bell-state analysis of two photons which are not brought together at the apparatus, so that Equation (\ref{conclus1}) can be rewritten in the form: 

\begin{equation}
N_{max, class}=N_{ZPF, A}-N_{ic}=N_{ZPF, A}-\frac{N_{ZPF, A}^{noise}}{2},
\label{conclus2}
\end{equation}
where, in Equation (\ref{conclus1}), we have replaced $N_{ZPF, S}$ by $N_{ZPF, A}$. Equations (\ref{conclus1}) and (\ref{conclus2}) are completely equivalent for experiments of BSM of two photons \cite{Walborn}, but the analysis below will be based in Equation (\ref{conclus2}). 

Let us now look at the Rome teleportation experiment. Apparently there is no relationship between the BSM in this experiment, in which the Bell-state analysis concerns only one of the two entangled photons, and the BSM of two photons. Nevertheless, the following similarities imply that Equation (\ref{conclus2}) can be used in the BSM performed in the Rome teleportation experiment:
\begin{enumerate}[(i)]
\item The two photons do not interact, i.e. they are not brought together at the experimental setup. This avoids the problems related to the bosonic symmetry of the photons \cite{B4}.
\item The use of enlarged Hilbert spaces (hyperentanglement) is a necessary step for complete BSM. In the WRHP formalism this is related to the inclusion of more sets of vacuum modes entering the source, which allows for the possibility of a positive balance between the ZPF sets of amplified modes entering the analyser, and the idle channels inside the analyser.
\item In such situation, the role of the ZPF inputs in the BSM of two non-interacting photons entangled in one degree of freedom is similar to the corresponding to one photon entangled in two independent degrees of freedom.
\end{enumerate}

In the Rome experiment, $N_{ZPF, S}=8$ is the number of ZPF sets of modes that are activated at the source of two-photon momentum entanglement,  and $N_{ZPF, A}=6$ is the number of ZPF sets of amplified modes entering the one-photon polarization-momentum Bell-state analyser, so that $N_{ZPF, S}\neq N_{ZPF, A}$. On the other hand, the number of entry points of noise inside the analyser is $N_{ic}=2$, in such a way that $N_{ZPF, A}^{noise}=4$. By applying Equation (\ref{conclus2}), we can determine the maximun number of one-photon polarization-momentum Bell-state classes that can be distinguished at the analyser: 
\begin{equation}
N_{max, class}=N_{ZPF, A}-\frac{N_{ZPF, A}^{noise}}{2}=6-2=4,
\label{conclus3}
\end{equation}
and this number will coincide with the number of Bell base states of one photon entangled in two degrees of freedom.

A similar scheme to the one proposed by Popescu was implemented by starting from polarization entanglement and  encoding the qubit to be teleported into the momentum \cite{Risco} of one of the entangled photons. To make this possible the Preparer uses a beam-splitter and phase shifters in one of the two beams. The zeropoint field entering the idle channel of the beam-splitter adds two sets of vacuum modes to the four sets provided by the source, so that the two beams entering the analyser contain six sets of vacuum modes. In this case, $N_{ZPF, S}=4$, $N_{ZPF, A}=6$, $N_{ic}=2$, and $N_{ZPF, A}^{noise}=4$, so that Equation (\ref{conclus2}) is fulfilled, where $N_{max, class}=6-2=4$. 

The two bits of classical communication that Alice sends to Bob constitute a key aspect in teleportation. The four qubits given in Equations (\ref{nueva}) or (\ref{pura}) are not orthogonal on each other, and therefore they cannot be measured with the same device. Hence, the apparatuses placed at the verification
station can only measure with certainty the two orthogonal states corresponding to a given basis.  As a consequence, the verification needs the classical information from Alice in order to generate the
graphs of Figures 2 and 3 of reference \cite{14}. 

The WRHP approach gives an explanation of this situation in terms of the number of the ZPF inputs at Bob's station. As we have demonstrated in Equation (\ref{conclus3}), the balance between the zeropoint inputs at Alice's station is adequate for measuring the four one-photon polarization-momentum Bell-states, but this property is not symmetric with respect to Bob. The balance between the amplified sets of ZPF modes entering Bob's station, in which the quantum information is stored, and the total number of ZPF inputs inside the verification station, is not sufficient to measure the four qubits given in Equation (\ref{pura}):
\begin{itemize}
\item On the one hand, in order to transform the cross-correlation properties concerning the momentum of Bob's photon (see Equations (\ref{v}) to (\ref{u})) into the cross-correla\-tions involving the polarization (see Equations (\ref{ok1}) to (\ref{44})), there must be a transfer from momentum to polarization. The excess of noise at Bob's station is eliminated through one of the outgoing channels of $\rm PBS_{Bob}$ (see Equation (\ref{noise})). This is just the contribution of the horizontal polarization components of the ZPF beams entering PBSs 1 and 2. Because of the beams entering Bob's station contain information concerning six sets of vacuum modes, once $F_{ZPF1, H}^{(+)}$ and $F_{ZPF2, H}^{(+)}$ are eliminated, 6-2=4 sets of amplified ZPF modes remain inside Bob's station: each polarization component of the beam ${\bf F}_{Bob,signal}^{(+)}$ (see Equation (\ref{40})) stores information of two ZPF sets of modes entering the crystal. 
\item By taking into consideration that there is a vacuum input channel at $ \rm PBS_{vs}$, the difference $4-1=3<4$ reveals the impossibility of distinghishing the four states. At this stage, the two bits that Alice sends to Bob contain the necessary classical information that completes the insufficient information at Bob's side. In other words, the vacuum input channel inside the verification station is compensated by means of two bits of classical information coming from Alice.
\end{itemize}
\section{Conclusions}\label{sec5}
In this paper, we have applied the WRHP approach of quantum optics to the study of the Rome teleportation experiment. We have investigated the physical meaning of the ZPF inputs at the different areas of the experimental setup.

In the WRHP formalism, entanglement appears just as an interplay of correlated waves, through the
essential contribution of the vacuum zeropoint field. The zeropoint entries contribute to storing quantum information into the field amplitudes at the source of entanglement. With an increasing number of vacuum inputs at the source, the possibility of extracting more information from the zeropoint field also increases, and this is a key aspect in $\rm 100\%$ success teleportation. The transfer from polarization entanglement to momentum entanglement supposes an increase of the activated ZPF sets of modes, from four to eight. In consequence, the Preparer can store information into the vacuum amplitudes $F_{ZPF1,V}^{(+)}$ and $ F_{ZPF2,V}^{(+)}$, which leads to an increase of the sets of amplified modes entering the analyser.

The ZPF inputs corresponding to the idle channels inside Alice's station contribute to the extraction of the two bits of classical information through the measurement process \cite{Hy}, in which the zeropoint intensity is subtracted at the detectors. These entry point of noise reduce the maximal classical information that can be extracted, in relation to the number of amplified sets of ZPF modes entering the  analyser. In this paper, we have demonstrated that the validity of Equation (36) of reference \cite{Hy} goes beyond the experimental situations concerning the Bell-state analysis of two photons that do not interact at the analyser, and it can be applied to the present situation in which the Bell-states concerning two independent degrees of freedom of only one of the two photons are measured. The maximal distinguishability is equal to the difference between the number of amplified ZPF sets of modes entering the analyser, and the number of idle channels (entry points of noise) inside the apparatus.  

Following the standard particle-like Hilbert space approach, the application of the projection postulate implies that the quantum information that remains in the electromagnetic field after the BSM is stored at Bob's station into a given one-photon polarization state, depending on the result obtained at Alice's station (see Equation (\ref{pura})). In contrast, the autocorrelations of the light field corresponding to ${\bf F}_{Bob,signal}^{(+)}$ do not give any information about the preparation (see Equations (\ref{40}) and (\ref{joycor})). Hence, the properties concerning Bob's photon after the BSM are supported by the cross-correlations associated to the field (see Equations (\ref{ok1}) to (\ref{44})): the behaviour of Bob's photon is inherently linked to Alice's photon, so that there is no disconnection between them mediated through the collapse of the state vector at Alice's station. The information that the Preparer includes into the electromagnetic field is conserved after the detection at Alice's station, and can be used through the photon at Bob's side after the classical communication. Nevertheless, the physical properties of Bob's photon must be deduced from the cross-correlation properties, where  the space-time condition given in Equation (\ref{coherence}) must be fulfilled. 

Also, the asymmetry between Alice and Bob in relation of the number of ZPF inputs has been analysed. The classical communication between Alice and Bob can be seen as a way to compensate the insufficient balance between the amplified ZPF sets of modes entering the verification station and the idle channel of $\rm PBS_{vs}$. A further study of the relationship between ZPF inputs and the trade-off between entanglement and classical-communication cost would be an interesting subject.

An established criteria for teleportation is that the unknown quantum state must come from outside \cite{A1}. This is not fulfilled in the Rome experiment, where the qubit to be teleported is encoded in one of two independent degrees of freedom of Alice's photon. Regardless of this issue, the WRHP formalism adds, in our opinion, a new perspective on the role and importance of the Rome teleportation experiment: the situations concerning the BSM of two photons that are not mixed at the device \cite{Walborn}, and the BSM of one-photon belonging to an entangled pair of non-interacting photons, converge though Equation (\ref{conclus2}). Hence, the possibility of performing a complete BSM in this experiment is not merely a way to demonstrate some elegant properties of the Hilbert-space formalism, but the most efficient use of the possibilities offered by the ZPF fluctuations in the absence of two-photon interaction at the analyser.

The Wigner representation, being an equivalent approach to the standard Hilbert space formalism, emphasizes the role of the zeropoint inputs, and contributes to a better understanding of quantum communication with light. The zeropoint field has a visible presence in the experiments, and this is just what the
WRHP formalism does: to open the possibility for exploring the physical
influence of vacuum fluctuations, not just through the commutation relations
between creation and annihilation mode operators, and the use of normal
ordering operators in photodetection, but seeking the physical meaning of
the zeropoint inputs at each step of a concrete experiment.

\newpage

\appendix
\numberwithin{equation}{section}

\section{Appendix A: General aspects of the WR\-HP formalism}
\label{A}
In this appendix, a brief review of the WRHP approach of PDC is
provided. 
The Wigner representation in the Heisenberg picture establishes a correspondence between the electric field operator and a time-dependent complex amplitude of the field, being the Wigner function time-independent. 
In the context of PDC the initial state is the vacuum, being the Wigner
distribution for the vacuum field amplitudes the Gaussian \cite{PDC7}:

\begin{equation}
W_{\it ZPF}(\{\alpha\})=
{\prod_{{\bf [k]}, \lambda}}\frac{2}{\pi}
{\rm e}^{-2|\alpha_{{\bf k}, \lambda}|^2},
\label{eq_w9}
\end{equation}
where $\alpha_{{\bf k}, \lambda}$ represents the zeropoint amplitude
corresponding to the mode $\{{\bf k}, \lambda\}$, and $\{\alpha\}$
represents the set of zeropoint amplitudes. The electric field corresponding to a signal beam generated by the
nonlinear crystal is represented by a slowly
varying amplitude \cite{pdc4}:

\begin{equation}
{\bf F}^{(+)}({\bf r}, t)=i{\rm e}^{\omega_s
t}\sum_{{\bf k}\in [{\bf k}]_{s}, \lambda=H, V}\left(\frac{\hbar
\omega_{{\bf k}}}
{2\epsilon_0L^3}\right)^{\frac{1}{2}}\alpha_{{\bf k}, \lambda}(0){\bf
u}_{{\bf k}, \lambda}{\rm e}^{i({\bf k}\cdot{\bf r}-\omega_{\bf k}t)},
\label{F}
\end{equation}
where $[{\bf k}]_{s}$ represents a set of wave vectors centred at
${\bf k}_s$, $\omega_s$ is the average frequency of the beam, and ${\bf
u}_{{\bf k}, \lambda}$ is a unit polarization vector. The amplitude $\alpha_{{\bf k}, \lambda}(0)$ is a linear transformation, to
second order in the coupling constant ($g$) of the zeropoint field
entering the nonlinear crystal, which interacts with the laser beam
between $t=-\Delta t$ and $t=0$, $\Delta t$ being the interaction time.
For $t>0$ there is a free evolution.
The field amplitude ${\bf F}^{(+)}$ propagates through free space
according to the following expression \cite{PDC7}:

\begin{equation}
{\bf F}^{(+)}({\bf{r}}_{2},t)=
{\bf F}^{(+)}({\bf{r}}_{1}, t-\frac{r_{12}}{c}){\rm e}^{i\omega_{s}\frac{r_{12}}{c}}
\,\,\,\,;\,\,\,\, {\bf{r}}_{12}=|{\bf{r}}_2-{\bf{r}}_1|.
\label{propagation}
\end{equation}

Given two complex amplitudes, $A({\bf r}, t; \{\alpha\})$ and $B({\bf
r'}, t';\{\alpha\})$, the correlation between them is given by:

\begin{equation}
\langle AB \rangle \equiv \int W_{\it ZPF}(\{\alpha\})A({\bf r}, t; \{\alpha\})B({\bf r'}, t';
\{\alpha\})d\{\alpha\}.
\label{corr}
\end{equation}
From Equation (\ref{eq_w9}) the following correlation properties
hold:

\begin{equation}
\langle \alpha_{{\bf k}, \lambda}\alpha_{{\bf k'}, \lambda'}\rangle
=\langle \alpha^*_{{\bf k}, \lambda}\alpha^*_{{\bf k'},
\lambda'}\rangle=0\,\,\,\,;\,\,\,\,\,\langle \alpha_{{\bf k}, \lambda}\alpha^*_{{\bf k'}, \lambda'}\rangle
=\frac{1}{2}\delta_{{\bf k}, {\bf k'}}\delta_{\lambda, \lambda'},
\label{correlations}
\end{equation}
which constitute an essential ingredient in order to describe entanglement in the WRHP approach.

The single and joint detection probabilities in PDC experiments are
calculated, in the Wigner approach, by means of the expressions
\cite{pdc4}:

\begin{equation}
P_{a}({\bf r}_{a},t)\propto \left\langle  I({\bf r}_{a},t)-I_{ZPF}({\bf r}_{a}) \right\rangle,
\label{probsimple}
\end{equation}
\begin{equation}
P_{ab}({\bf r}_{a},t;{\bf r}_{b},t')\propto \left\langle  [I({\bf r}_{a},t)-I_{ZPF}({\bf r}_{a})] [I({\bf r}_{b},t')-I_{ZPF}({\bf r}_{b})]\right\rangle,
\label{prob}
\end{equation}
where $I({\bf r}_{i},t)\propto |{\bf F}^{(+)}({\bf r}_{i},t)|^{2}$, $i=\{a, b\}$, is the
intensity of light at the position of the $i$-detector, and $I_{ZPF}({\bf r}_{i})$ is the corresponding intensity of the zeropoint field. Equation (\ref{prob}) is fulfilled in the case that the field operators corresponding to detectors $a$ and $b$ commute, which is the situation in an important part of the experiments performed using PDC. In experiments involving polarization, the following simplified
expression for the joint detection probability is used for
practical matters:

\begin{equation}
P_{ab} \left({\bf{r}},t;{\bf{r'}},t'\right)\propto
\sum _{\lambda, \lambda'}
\left|\left\langle F_{a, {\lambda }}^{\left(+\right)} \left(\phi_{a} ;{\bf{r}},t\right)
F_{b, {\lambda '}}^{\left(+\right)} \left(\phi_{b} ;{\bf{r'}},
t'\right)\right\rangle \right|^{2},
\label{p12}
\end{equation}
where $\phi_{a}$ and $\phi_{b}$ represent controllable parameters of the
experimental setup.

In actual experiments, the expressions (\ref{probsimple}) to (\ref{p12}) must be integrated over the surface of the detectors and appropriate detection windows.

A key point of the WRHP formalism of PDC is the description of polarization entanglement. This property appears just as an
interplay of correlated waves, through the distribution of the vacuum
amplitudes in the different polarization components of the field
\cite{pdc4}. For instance, the quantum predictions corresponding to the polarization
state $|\Psi^{+}\rangle=(1/\sqrt{2})(|H\rangle_{1}|V\rangle_{2}+|V\rangle_{1}|H\rangle_{2})$ are reproduced in the Wigner framework by
considering the following two correlated beams going out of the crystal
\cite{CD}, which are generated in the Rome teleportation experiment:
\begin{equation}
{\bf F}_{1}^{(+)}({\bf r}, t)=F_s^{(+)}({\bf r},
t; \{\alpha_{{\bf k}_1, H}; \alpha^*_{{\bf k}_2, V}\}){\bf
i}_{1}+F_{p}^{(+)}({\bf r}, t;
\{\alpha_{{\bf k}_1, V}; \alpha^*_{{\bf k}_2, H}\}){\bf j}_{1}
\label{hyper0},
\end{equation}
\begin{equation}
{\bf F}_{2}^{(+)}({\bf r}, t)=F_{q}^{(+)}({\bf r},
t; \{\alpha_{{\bf k}_2, H}; \alpha^*_{{\bf k}_1, V}\}){\bf
i}_{2}+ F_{r}^{(+)}({\bf r}, t;
\{\alpha_{{\bf k}_2, V}; \alpha^*_{{\bf k}_1, H}\}){\bf j}_{2},
\label{hyper3}
\end{equation}
where ${\bf i}_{1}$ and ${\bf i}_{2}$ (${\bf j}_{1}$ and ${\bf j}_{2}$) are unit
vectors representing horizontal (vertical) linear polarization at beams
``$1$" and ``$2$", and $\{\alpha_{{\bf k}_i, V};\alpha_{{\bf k}_i,
H}\}$\,($i=1, 2$) represent four sets of relevant zeropoint amplitudes
entering the crystal. The four set of modes $\{{\bf k}_{i,
\lambda}\}$\,($i=1, 2$;  $\lambda\equiv H, V$) are ``activated" throughout the coupling with the laser beam inside the nonlinear medium. In expressions
(\ref{hyper0}) and (\ref{hyper3}) the only non vanishing cross-correlations
are those involving the combinations
$r\leftrightarrow s$ and $p\leftrightarrow q$. Hence, the non-null cross-correlations
correspond to different polarization components, which is a consequence of Equation (\ref{correlations}).

By using Equation (\ref{propagation}), the cross-correlations, at any
position and time, can be expressed in terms of the corresponding ones
at the centre of the nonlinear source \cite{pdc4}. We have:
\begin{equation}
\langle F^{(+)}_{r}({\bf{r}}_{C},
t)F^{(+)}_{s}({\bf{r}}_{C}, t')\rangle=\langle F^{(+)}_{p}({\bf{r}}_{C}, t)F^{(+)}_{q}({\bf{r}}_{C}, t')\rangle =gV\nu(t'-t),
\label{nu}
\end{equation}
where $V$ is the amplitude of the laser beam. $\nu(t'-t)$ is a function
which vanishes when $|t'-t|$ is greater than the correlation  time  between the amplitudes $F^{(+)}_{p}$ ($F^{(+)}_{r}$) and $F^{(+)}_{q}$ ($F^{(+)}_{s}$) \cite{PDC8}.

On the other hand, by considering the amplitud $F_{s}^{(+)}$ at the position ${\bf{r}}$ and times $t$ and $t'$, the following autocorrelation property holds: 
\begin{equation}
\langle F^{(+)}_{s}({\bf{r}}, t)F^{(-)}_{s}({\bf{r}}, t')\rangle-\langle [{\bf F}^{(+)}_{ZPF,1}({\bf{r}}, t)\cdot {\bf{i}}_{1}] [{\bf F}^{(-)}_{ZPF,1}({\bf{r}}, t')\cdot {\bf{i}}_{1}]\rangle =\frac{g^{2}|V|^{2}}{2}\mu(t'-t), 
\label{new}
\end{equation}
where ${\bf F}^{(+)}_{ZPF,1}$ is the zeropoint beam corresponding to mode ${\bf{k}}_{1}$, and $\mu(t'-t)$ is a correlation function which goes to zero when $|t'-t|$ is greater than the coherence time of PDC light. Similar expressions hold for $F^{(+)}_{p}$, $F^{(+)}_{q}$ and $F^{(+)}_{r}$.

The rest of the polarization Bell-states have been analysed in the WRHP formalism \cite{CD} along with the description of polarization-momentum hyperentanglement of two photons \cite{Hy}.
\section{Appendix B: Popescu's protocol for teleportation experiment in the
Hilbert space}\label{B}
In this appendix we shall briefly describe the basic aspects of Popescu's scheme for teleportation in the Hilbert space \cite{13, ekert}. The concepts and equations presented in this appendix are used throughout this paper, in order to emphasize the differences with the WRHP description.

The first step is to produce two entangled photons in momentum, but each of them with well defined polarization. To this end, an entangled state in polarization is generated using type-II parametric down conversion:
\begin{equation}\begin{array}{l}{| \Psi ^{(+)}  \rangle} =\frac{1}{\sqrt{2} } [{| H \rangle} _{1} {| V \rangle} _{2} +{| V \rangle} {_{1} } {| H \rangle} _{2} ]{| a_{1}  \rangle}{| b_{2}  \rangle},\label{Hilbert1}\end{array}\end{equation} 
where 1 and 2 label the two output directions of the entangled photons. In order to transfer the polarization entanglement to momentum entanglement, each beam passes through a polarizing beamsplitter (PBS) that transmits (reflects) vertical (horizontal) polarization. The resulting two-photon state can be expressed as:

\begin{equation}\begin{array}{l}{| \Phi_{in}  \rangle} =\frac{i}{\sqrt{2} } ({| a_{1}  \rangle} {| a_{2}  \rangle} +{| b_{1}  \rangle} {| b_{2}  \rangle} ){| H \rangle} _{1} {| V \rangle} _{2},\label{Hilbert2} \end{array}\end{equation}  
where index 1 (horizontal polarization) and index 2 (vertical polarization) now refer to the photon directed to Alice's  station and Bob's station respectively. The unit imaginary number that appears in Equation (\ref{Hilbert2}) is related to the consideration of the factor $i$ in the reflection, and it will be considered through the different optical devices (see Figure \ref{Figure1}). The two-photon state given by Equation (\ref{Hilbert2}) constitutes the quantum-communication channel (see Figure \ref{Figure1}).

The second step consists in the  preparation of a generic state to be teleported. To this end, Alice photon is intercepted by the Preparer P, who acts in the same manner in both paths $a_{1}$ and $b_{1}$, by changing the polarization state  ${| H \rangle} _{1}$ to an arbitrary qubit:
\begin{equation}\begin{array}{l}{| \psi \rangle} _{1} =\alpha {| H \rangle} _{1} +\beta {| V \rangle} _{1} ,\label{Hilbert3}\end{array}\end{equation}
so that the two-photon state after the preparation is:

\begin{equation}\begin{array}{l}{| \Phi_{P} \rangle} =\frac{i}{\sqrt{2} } ({| a_{1} \rangle} {| a_{2}  \rangle} +{| b_{1}  \rangle} {| b_{2}  \rangle} ){| \psi \rangle} _{1} {| V \rangle} _{2} .\label{Hilbert4}\end{array}\end{equation} 

The state given in Equation (\ref{Hilbert4}) is formally  equivalent to the following superposition in the four Bell states representing entanglement between polarization and momentum of photon $1$:

\[{| \Phi_{P} \rangle} =\frac{i}{2} {| c_{+} \rangle} (\beta {| a_{2} \rangle} +\alpha {| b_{2} \rangle} ){| V \rangle} _{2} +\frac{i}{2} {| c_{-} \rangle} (\beta {| a_{2}\rangle} -\alpha {| b_{2}  \rangle} ){| V \rangle} _{2} \] 
\begin{equation}\begin{array}{l}
+\frac{i}{2} {| d_{+} \rangle} (\alpha {| a_{2}  \rangle} +\beta {| b_{2} \rangle} ){| V \rangle} _{2} +\frac{i}{2} {| d_{-}  \rangle} (\alpha {| a_{2} \rangle} -\beta {| b_{2} \rangle} ){| V \rangle} _{2},\label{nueva} \end{array}\end{equation} 
where: 
\begin{equation}\begin{array}{l}{| c_{\pm }  \rangle} =\frac{1}{\sqrt{2} } ({| a_{1}  \rangle} {| V \rangle} _{1} \pm {| b_{1}  \rangle} {| H \rangle} _{1} ),\end{array}\end{equation} 
\begin{equation}\begin{array}{l}{| d_{\pm } \rangle} =\frac{1}{\sqrt{2} } ({| a_{1} \rangle} {| H \rangle} _{1} \pm {| b_{1} \rangle} {| V \rangle} _{1} ).\end{array}\end{equation} 
In the third step of the teleportation protocol, Alice performs a BSM on the basis $\{{| c_{\pm } \rangle},{| d_{\pm } \rangle}\}$, for which the polarization and momentum of photon $1$ have to be entangled. The interaction between the polarization and direction is achieved by using PBSs on the paths $a_{1}$ and $b_{1}$, so that the information about polarization is encoded in the position of the photon. By using polarization rotators ($90^{\circ}$) and balanced beam-splitters, a photon detection in one of the four detectors $\rm DX\pm$ ($X=T,R$) will correspond to a projection onto one of the four Bell-states. The state of the two photons just before the detection of photon $1$ by  one of the detectors $\rm DX\pm$ ($X=\{T, R\}$), and by taking into consideration that Bob transforms the superposition in momentum of photon $2$ into same superposition in polarization, is:
\[|\Phi_{det}\rangle=\frac{1}{2} \{{i| DT- \rangle} (\alpha {| H \rangle}_{2} + \beta {| V \rangle} _{2}) - {| DT+ \rangle} (\alpha {| H\rangle}_{2} -\beta {|V  \rangle}_{2} ) \] 
\begin{equation}\begin{array}{l}
+ {| DR- \rangle} (\beta {| H \rangle}_{2} +\alpha {| V \rangle}_{2} )-i{| DR+  \rangle} (\beta {|H \rangle}_{2} -\alpha {|V \rangle}_{2} )\}|H \rangle_{1} |S_{B} \rangle,\label{pura} \end{array}\end{equation} 
where ${| DX\pm \rangle}$ ($X=\{T, R\}$) represent the momentum states that are directly detected by detectors $\rm DX\pm$, and ${| S_{B}\rangle}$ denotes the momentum state corresponding to Bob's signal. 

The fourth step of the protocol comes when Alice informs  Bob (through a classical communication channel) which detector registers a photon. With this information, Bob can reproduce the superposition given in Equation (\ref{Hilbert3}). He uses two optical elements $C_{1}$ and $C_{2}$ in order to transform the polarization state of photon 2 into the prepared superposition.
\newpage

\newpage
\begin{thebibliography}{99}

\bibitem{In3}Bennett, C. H.; Brassard, G.; Cr\'epeau, C.; Jozsa, R.; Peres, A.;
Wootters, W. K. \textit{Phys. Rev. Lett.} {\bf{1993}}, \textit{70}, 1895-1899.
\bibitem{A1} Pirandola, S.; Eisert, J.;	Weedbrook, C.; Furusawa, A.; Braunstein, S. L. \textit{Nat. Photon.} {\bf{2015}}, \textit{9}, 641-652.
\bibitem{tr4}Metcalf, B. J. \textit{et al.} \textit{Nat. Photon.} {\bf{2014}}, {\textit{8}}, 770-774.
\bibitem{tr5}Kiktenko, E. O.; Popov, A. A.; Fedorov, A. K. \textit{Phys. Rev. A} {\bf{2016}}, {\textit{93}}, 062305, 8 pages.
\bibitem{Graham} Graham, T. M.; Bernstein, H. J.; Wei, T.-C.; Junge, M.; Kwiat, P. G. \textit{Nat. Commu.} {\bf 2015}, \textit{6}, 7185, 9 pages. DOI: 10.1038/ncomms8185. Published Online: May 28, 2015.
\bibitem{A}  Lo, H.-K.  \textit{Phys. Rev. A} {\bf{2000}},  \textit{62}, 012313, 7 pages.
\bibitem{B} Pati, A. K.  \textit{Phys. Rev. A} {\bf{2000}}, \textit{63}, 014302, 3 pages.
\bibitem{C}  Bennett, C. H. \textit{et al.} \textit{Phys. Rev. Lett.} {\bf{2001}}, \textit{87}, 077902, 4 pages.
\bibitem{B1}Franson, J. D. \textit{Phys. Rev. Lett.} {\bf{1989}}, \textit{62}, 2205-2208.
\bibitem{B2} Rarity, J. G.; Tapster, P. R.  \textit{Phys. Rev. Lett.} {\bf{1990}}, \textit{64}, 2495-2498.
\bibitem{B3} Ekert, A. K. \textit{Phys. Rev. Lett.} {\bf{1991}}, \textit{67}, 661-663.
\bibitem{B4}Mattle, K.; Weinfurter H.; Kwait, P. G.;  Zeilinger, A. \textit{Phys. Rev. Lett.} {\bf{1996}}, \textit{76}, 4656-4659.
\bibitem{B5} Ribordy, G.; Brendel, J.; Gautier, J. D.; Gisin, N.; Zbinden, H. \textit{Phys. Rev. Lett. A} {\bf{2001}}, \textit{63}, 012309, 12 pages.
\bibitem{PDC1} Bouwmeester, D.; Pan, J.-W.; Mattle, K.; Eilb, M.; Weinfurter, H.; Zeilinger, A. {\textit{Nature}} {\bf{1997}}, \textsl{390}, 575-579.
\bibitem{14} Boschi, D.; Branca, S.; De Martini, F.; Hardy, L.; Popescu, S. \textit{Phys. Rev. Lett.} {\bf{1998}}, \textit{80}, 1121-1125.
\bibitem{15} Jennewein, T.; Weihs, G.; Pan, J.W.; Zeilinger, A. \textit{Phys. Rev. Lett.} {\bf{2002}}, \textit{88}, 017903, 4 pages.
\bibitem{16} Lombardi, E.; Sciarrino, F.; Popescu, S.; De Martini, F. \textit{Phys. Rev. Lett.} {\bf{2002}}, \textit{88}, 070402, 4 pages.
\bibitem{17} Marcikic, I.; de Riedmatten, H.; Tittel, W.; Zbinden, H; Gisin, N. \textit{Nature} {\bf{2003}}, \textit{421}, 509-513.
 \bibitem{nueva}Kim Y.-H.; Kulik S.P.; Shih Y. \textit{Phys. Rev. Lett.} {\bf{2001}}, \textit{86}, 1370-1373.
 \bibitem{13} Popescu, S. {\bf{1995}}, LANL Archives quant-ph/9501020.
 \bibitem{Niel} Nielsen, M. A.; Knill, E.; Laflamme, R. \textit{Nature} {\bf 1998}, \textit{396}, 52-55.
 \bibitem{Ma} Ma, X.-S. et al. \textit{Nature} {\bf 2012}, \textit{489}, 269-273.
\bibitem{PDC7} Casado, A.; Fern\'andez-Rueda, A.; Marshall, T.; Risco-Delgado, R.; Santos, E. \textit{Phys. Rev. A} {\bf{1997}}, \textit{55}, 3879-3890.
\bibitem{pdc4} Casado, A.; Marshall, T.; Santos, E. \textit{J. Opt. Soc. Am. B} {\bf{1998}},
{\textit{15}}, 1572-1577.
\bibitem{E1} Santos, E. \textit{Proc. SPIE} {\bf{2005}}, 5866, 36-37. doi:10.1117/12.619611. Published
online: August 04, 2005.
\bibitem{PDC8} Casado, A.; Fern\'andez-Rueda, A.; Marshall, T.; Mart\'inez, J.; Risco-Delgado, R.; Santos, E. \textit{Eur. Phys. J. D} {\bf{2000}}, {\textit{11}}, 465-472.
 \bibitem{cr}Casado, A.; Guerra, S.; Pl\'acido, J. \textit{J. Phys. B: At. Mol. Opt.
Phys}. {\bf{2008}}, \textit{41}, 045501, 7 pages.
 \bibitem{CD}Casado, A.; Guerra, S.; Pl\'acido, J. \textit{Adv. Math. Phys.} {\bf{2010}},
\textit{2010}, 501521, 11 pages.
\bibitem{Hy} Casado, A.; Guerra, S.; Pl\'acido, J. \textit{Eur. Phys. J. D} {\bf{2014}}, \textit{68}, 338, 11 pages.
 \bibitem{sw}Casado, A.; Guerra, S.; Pl\'acido, J. \textit{Journal of Modern Optics} {\bf{2014}}, \textit{62}
377-386.
\bibitem{Pisenti} Pisenti, N.;  Gaebler, C.P.E.; Lynn T.W. \textit{Phys. Rev. A} {\bf{2011}}, \textit{84}, 022340, 5 pages.
\bibitem{ekert} Bouwmeester, D.; Ekert, A.K.; Zeilinger, A. \textit{Springer-Verlag}, Berlin, {\bf{2000}}.

\bibitem{Walborn} Walborn, S.P.; P\'adua, S.;  Monken, C.H. \textit{Phys. Rev. A} {\bf 2003}, \textit{68},
042313, 5 pages. 
\bibitem{Risco} {Michler, M.; Risco-Delgado, R.; Weinfurter, H.} Technical Digest, EQEC {\bf{1998}}, 1 page, doi: 10.1109/EQEC.1998.714868. Published online: August 06, 2002.

\end{thebibliography}
\end{document}